\RequirePackage{fix-cm}
\documentclass[smallextended]{svjour3}       
\smartqed  
\usepackage{graphicx}
\usepackage{color}
\begin{document}

\title{Statistics of particle accumulation in spatially developing turbulent boundary layers
}
\titlerunning{Particle accumulation in turbulent boundary layers}        

\author{G. Sardina  \and  F. Picano \and P. Schlatter  \and L. Brandt  \and C.M. Casciola}

\institute{
           G. Sardina \at
          Facolt\`a di Ingegneria, Architettura e Scienze Motorie, \\UKE Universit\`a Kore di Enna, 94100 Enna, Italy\\
              and Linn\'e FLOW Centre, 
              KTH Mechanics, SE-100 44 Stockholm, Sweden\\
              \email{gaetano@mech.kth.se}           
     \and
              F. Picano \and P. Schlatter  \and L. Brandt \at
              Linn\'e FLOW Centre and SeRC (Swedish e-Science Research  Centre)\\
              KTH Mechanics, SE-100 44 Stockholm, Sweden\\
           \and
           C. M. Casciola \at
             Dipartimento di Ingegneria Meccanica e Aerospaziale, Sapienza University of
	      Rome, \\via Eudossiana 18, 00184 Rome, Italy
}

\date{Received: date / Accepted: date}

\maketitle
\sloppy
\begin{abstract}
We present the results of a Direct Numerical Simulation of a particle-laden spatially developing
turbulent boundary layer up to $Re_\theta=2500$.
Two main features differentiate the behavior of inertial particles in a zero-pressure-gradient turbulent boundary 
layer from the more commonly studied case of a parallel channel flow. The first is
the variation along the streamwise direction of the local dimensionless parameters defining the fluid-particle
interactions. The second is the coexistence of an  irrotational free-stream and a near-wall rotational turbulent flow. 
As concerns the first issue, an inner and an outer Stokes number can be defined using inner and outer flow units.
The inner Stokes number governs the near-wall behavior similarly to the case of channel flow.
To understand the effect of a laminar-turbulent interface, we examine the behavior of particles initially released in the free stream  
and show that they present a distinct behavior with respect to those
directly injected inside the boundary layer. 
A region of minimum concentration occurs inside the turbulent boundary layer at about one displacement thickness from the wall.
Its formation is due to the competition between two transport mechanisms:  a relatively slow turbulent diffusion towards the buffer layer 
and a fast turbophoretic drift towards the wall.

\end{abstract}

\section{Introduction}

The dispersion of a solid phase in turbulent wall-bounded flows occurs
in many  technological processes such particle-blade interactions in the turbines of aeronautical engines.
Inertial particles transported in turbulent wall flows display a characteristic preferential accumulation close to the wall. This phenomenology
is denoted  turbophoresis~\cite{capetal,reeks} and has been the subject of research in the last three decades~\cite{baleat}.
The turbophoretic drift towards the wall is essentially controlled by one
nondimensional parameter, the viscous Stokes number, that is the ratio between the
particle relaxation time $\tau_p=\rho_p\, d_p^2/(18\, \rho\, \nu)$ and the viscous time scale of the flow $\nu/U_\tau^2$, with
$\rho_p$ and $d_p$ the particle density and diameter, $\rho$ and $\nu$ the fluid density and viscosity, and $U_\tau$ the
friction velocity.
In particular,
the strongest particle accumulation towards the wall is found when the Stokes
number is about twenty five, $St^+\sim25$. Particles of vanishing Stokes number behave as passive tracers and are therefore uniformly distributed in a turbulent flow, whereas heavy particles become insensitive to the turbulence fluctuations, the ballistic limit at high Stokes numbers.
A review of experimental studies and direct numerical simulation (DNS) of turbophoresis over the last years can be found in \cite{solrew}, with most recent simulations presented in \cite{sar_ftac,sar_jfm}.
In most of these previous investigations, the dynamics of the inertial particle has been
studied in parallel flows such as channels or pipes~\cite{sar_jfm,roueat,pic_pipe,marval}.
A different approach is necessary in spatially evolving flows where it is fundamental to understand the dynamics of the near-wall accumulation
when the local Stokes number of the dispersed phase changes during the
particle evolution, i.e.\ in the streamwise direction, leading to
non trivial effects  as those observed in particle-laden turbulent
round jets~\cite{pic_jet}.
In this context, we present here statistical data from a large-scale direct numerical simulations (DNS) of a spatially evolving
particle-laden turbulent boundary layer at Reynolds number $Re_\theta=2500$, based on the momentum thickness $\theta$
corresponding to a friction Reynolds number $Re_\tau=800$ based on the friction velocity. 
This can be seen as a moderate Reynolds numbers in experiments but it is certainly high for a fully resolved numerical simulation~\cite{sch_2}.
The study of particle dynamics in spatial developing boundary
layers assumes a fundamental role to advance our understanding in multiphase flows because
it represents the most ideal flow  where the Stokes number changes
along the mean stream direction.

\begin{figure}[t!]
\begin{center}
\includegraphics[width=0.5\textwidth]{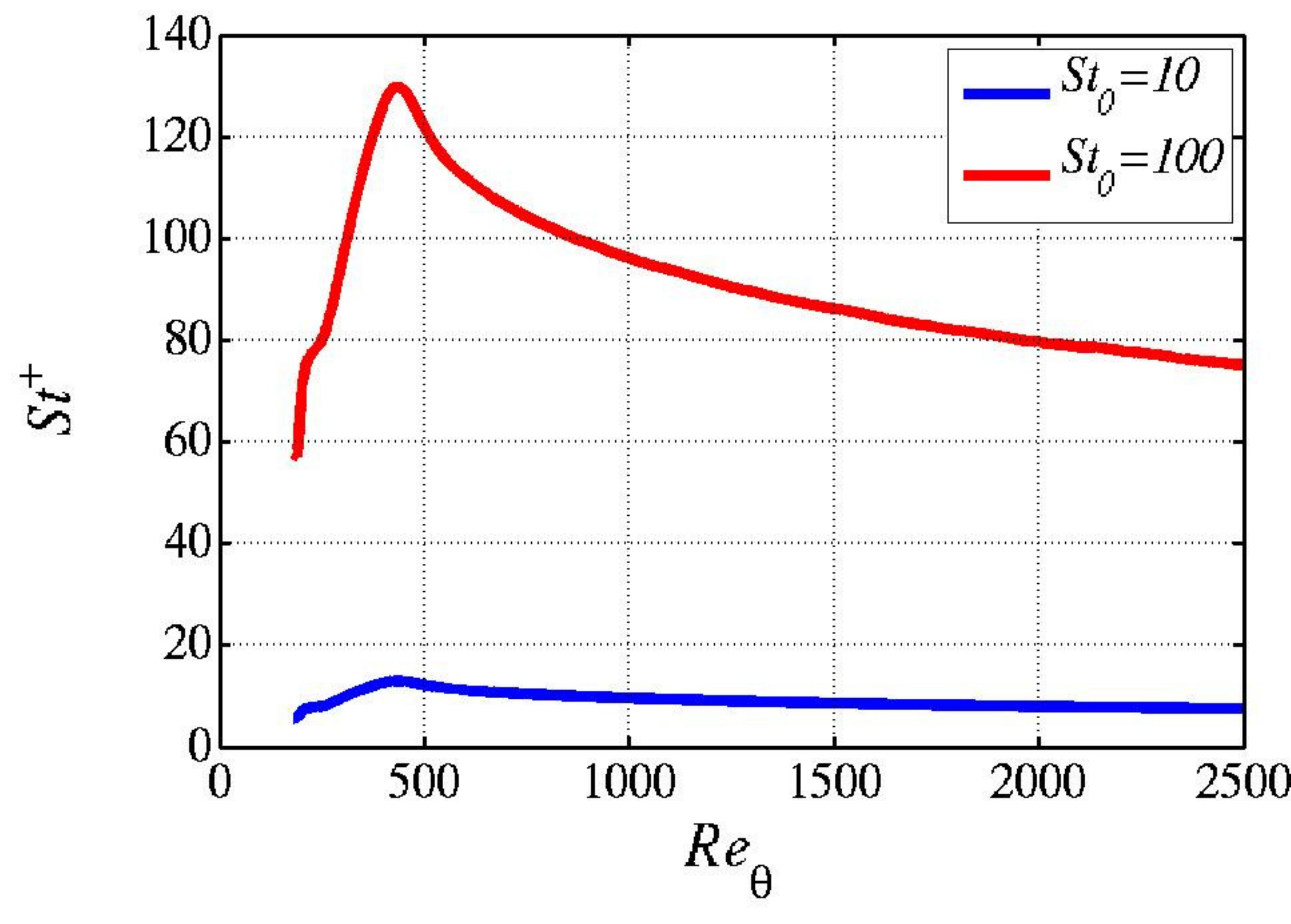}
\put(-130,120){$a)$}
\includegraphics[width=0.5\textwidth]{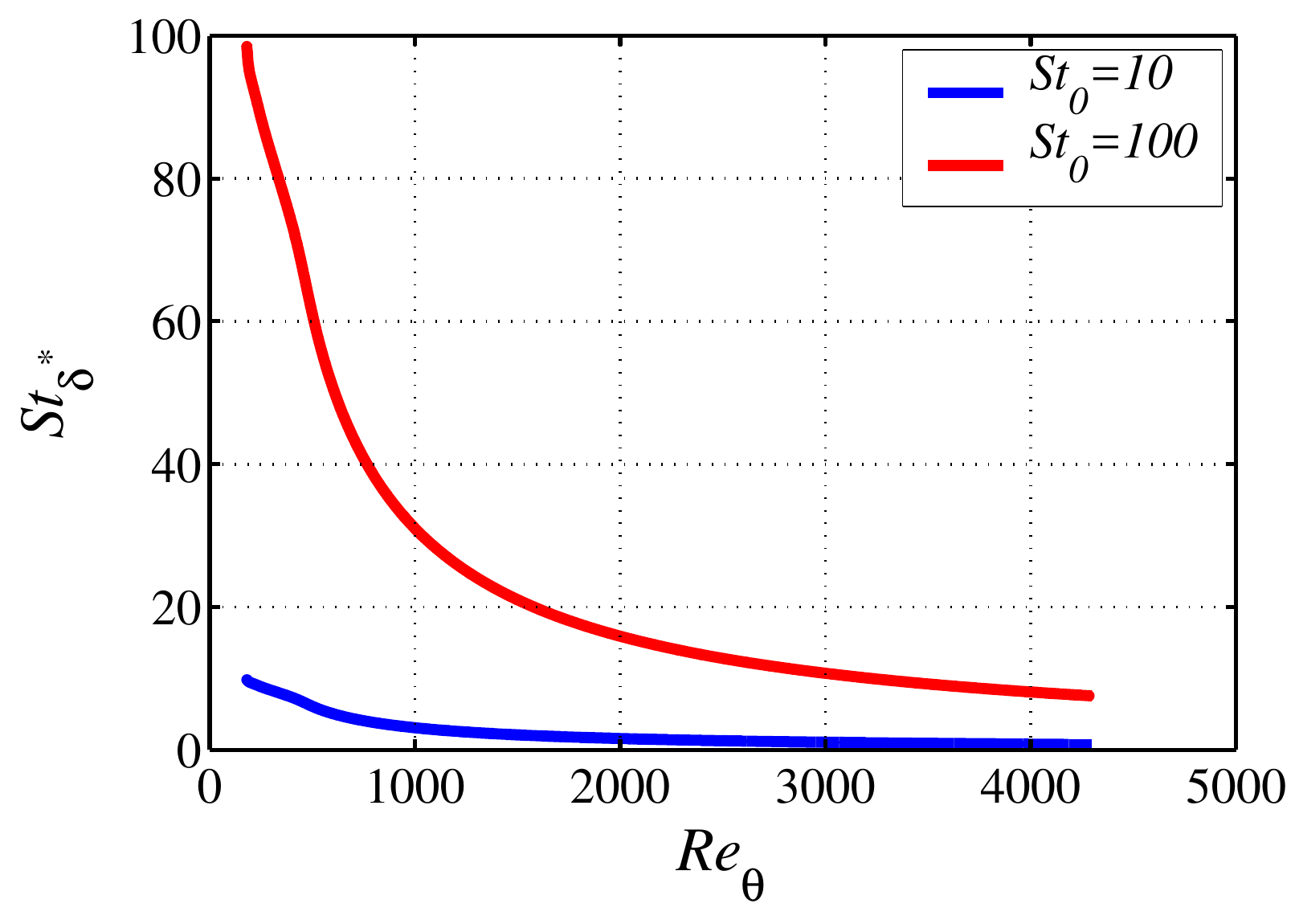}
\put(-130,120){$b)$}
\caption{$a)$ Development of the viscous Stokes number $St^+$ in streamwise
direction for two nominal Stokes numbers, $St_0$. $b)$ Development of the
external Stokes number $St_{\delta^*}$ in streamwise
direction for two nominal Stokes numbers, $St_0$. \label{fig1}}
\end{center}
\end{figure}

A nominal Stokes number $St_0=\tau_p\,U_\infty/\delta_0^*$ can be defined to measure the amount of inertia of a single particle population by using
the free-stream velocity $U_\infty$ and the displacement thickness of the boundary layer at the inflow $\delta_0^*$ of the computational domain (or any reference station such as the seeding location).
For a spatially evolving flow, more meaningful local Stokes numbers needs to
be defined, both based on local internal units $St^+$, as usually defined in parallel wall flows, and using external flow units $St_{\delta^*}$,
\begin{equation}
St^+=\frac{\tau_p U_\tau^2}{\nu}, \quad \quad \quad 
St_{\delta^*}=\frac{\tau_p U_\infty}{\delta^{*}},
\end{equation}
where it should be remarked that  the
displacement thickness of the boundary layer $\delta^*$ and the friction velocity $U_\tau$ change along the streamwise direction.
Indeed, the two Stokes numbers decrease moving downstream
in the boundary layer and tend to the limit of passive tracers. In figure~\ref{fig1} the two Stokes numbers are plotted versus $Re_\theta$, the Reynolds number based on the momentum thickness $\theta$, 
 usually adopted to parametrize the streamwise distance in turbulent boundary layers~\cite{sch_2}. 
As apparent by the decrease of the Stokes numbers, the effects of the inertia become less and less relevant in the downstream direction 
until the particles behave as  fluid tracers at sufficient distance from the leading edge. Note that the decay rate is different for the two parameters, something discussed in further detail below.
Both the inner and outer Stokes numbers are monotonically decreasing downstream when the boundary layer is fully turbulent; the peaks displayed by $St^+$ around $Re_\theta=400$ occur in the region where the transition from laminar to turbulent state takes place. 

Another peculiar aspect of the turbulent boundary layer is represented by the co-existence of two different zones, the inner turbulent region
and the external free stream. These two regions are separated by an intermittent interface, the viscous super-layer~\cite{corrsin},  
with a fractal nature where the enstrophy generated at the wall diffuses towards the outer irrotational free stream. 
In many applications, the particles usually lie in the outer region and then enter the turbulent inner region, crossing the separating interface.

This characteristic property, together with the spatial development of the turbulent boundary layer, introduces new features in the particle transport that
cannot be investigated in parallel wall flows. 
In our recent study~\cite{sardina2012self}, we examined the same set of simulations 
and show that the concentration and the streamwise
velocity profiles are self-similar and depend only on the local value
of the outer Stokes number and the rescaled wall-normal distance.

The aim of the present investigation is to further study the particle dynamics in a spatial boundary layer with emphasis on the distinction between particles that are seeded inside the turbulent regions and particles initially located outside the turbulent region, penetrating the shear layer and accumulating at the wall. 
In particular, we show that a minimum in the concentration profiles occurs at around one displacement thickness from the wall and this is the result
of the competition of two transport mechanisms both directed towards the wall, but of different intensities. 
Particles are first subjected to a slow dispersion process from the outer region to the buffer layer and then to a fast
turbophoretic drift close to the wall.

\section{Numerical method}

The numerical solver employed for the simulation is the pseudo-spectral code
SIMSON \cite{simson}.
The dimensions of the computational domain are $x_L\times y_L\times z_L=3000\delta_0^*\times 100 \delta_0^*\times 120 \delta_0^*$ in the streamwise, wall-normal
 and spanwise directions with $ \delta_0^* $ the displacement thickness at the 
inlet. The solution is expressed in terms of Fourier modes in the streamwise and 
spanwise direction with $n_x=4096$ and $n_z=384$, while  
 $n_y=301$ Chebyshev modes are used to discretize the wall-normal direction.  
Flow periodicity in the streamwise direction is handled with a fringe region at the end of the computational domain \cite{fringe} 
where the velocity field is forced to the laminar Blasius profile at $Re_{\delta_0^*}=400$.
The flow is tripped just downstream of the inlet by a localized forcing
random in time and in the spanwise direction to trigger the laminar-turbulent
transition. To reach a fully developed turbulent flow, the
carrier phase needs to reach a Reynolds
number of the order $Re_\theta\simeq 1600$. In the present case, $Re_\theta$ varies 
from $Re_\theta \simeq 200$ at the inflow to $Re_\theta = 2500 $ at the
end of the computational domain. The unladen reference simulation is described in \cite{sch_2} where the
same geometry and flow parameters are employed.

Regarding the dispersed phase,
we assume that the particle concentration is dilute and neglect the backreaction
on the flow, inter-particle collisions and hydrodynamic interactions among particles so that 
the one-way coupling approximation can be safely adopted.
The particles are assumed to be small, rigid spheres with density one
thousand times that of the carrier phase. The only force acting on the 
individual particle is the Stokes drag \cite{maxril}.
The fluid velocity at the particle position is evaluated by
 a fourth order spatial interpolation and  the particle time integration is performed with a
second order Adams-Bashforth scheme.
Seven populations, differing only in the nominal
Stokes number $St_0$, are evolved inside the computational domain.
Particles are injected at a constant rate into the already turbulent flow at
the streamwise location corresponding to
$Re_\theta\simeq 800$ so that their evolution is not directly influenced by the trip
forcing. For each population, particles are injected
at the rate of $64$ in $\delta_0^*/U_\infty$,   
randomly in the spanwise direction and at locations equispaced in the wall-normal direction in the range $y\in[0,4\delta_{99}]$.
Hence, a part of the total particles are released inside the turbulent region at a wall-normal distance below the boundary layer thickness
$\delta_{99}$, whereas the remaining particles are injected further away from the wall in the irrotational free stream. 
 An example of a corresponding physical case is a flat plate moving with velocity $U_\infty=5\, m s^{-1}$ in dusty
 air with particle diameters  $d_p=6\div60 \mu m$ and density ratio $\rho_p/\rho_f=1000$. 
 The computational domain describes the evolution of the particle-laden turbulent boundary
 layer until a momentum thickness $\theta=5 {\rm mm}$.    
Computational time has been provided within the DEISA project WALLPART.
Further details about the numerical simulation can be found in \cite{sardina2012self}.

\section{Results}

\subsection{Instantaneous visualizations}
\begin{figure}[tb!]
\begin{center}
\includegraphics[width=0.45\textwidth]{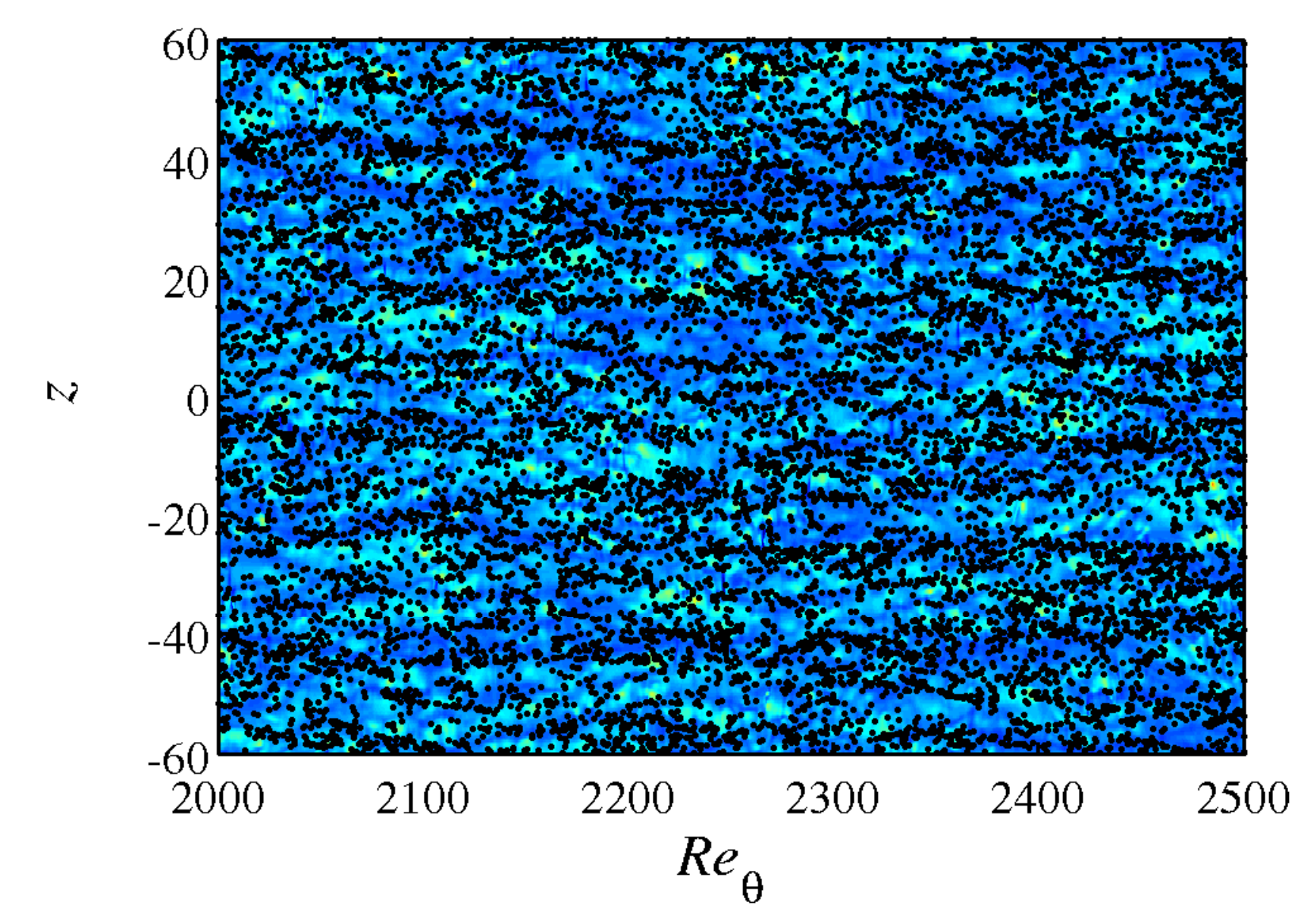}~~~
\put(-120,110){$a)$}
\includegraphics[width=0.45\textwidth]{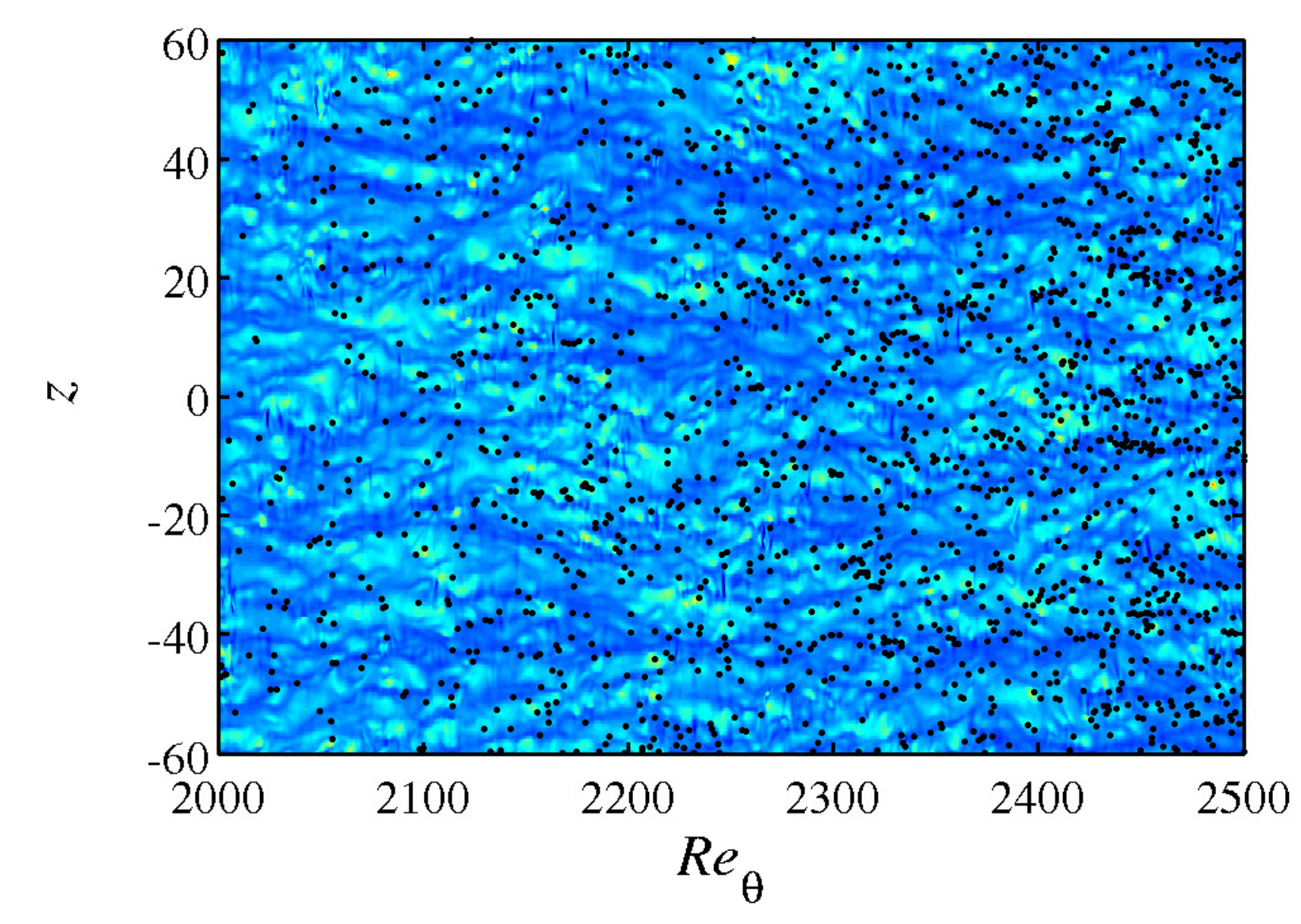}
\put(-120,110){$b)$}
\caption{Istantaneous configurations of inertial particles with nominal $St_0=30$ in 
the region close to the wall along a wall-parallel plane towards the end of the
computational domain. Colours represent the values of the wall-shear stress (higher values lighter zones).
 $a)$ Total particles in the given configuration. $b)$ 
Particles initially released at a wall-normal distance larger than the local boundary layer thickness.
 \label{fig2}}
\end{center}
\end{figure}

Two instantaneous configurations of the particle positions
close to the wall are shown in figure \ref{fig2} for the population with a nominal
Stokes number $St_0=30$ in the region close to the end of the domain, i.e.\ with
$Re_\theta$ ranging from 2000 to 2500.
The background color represents the values of the local wall-shear stress $\tau_w$, 
representative of the sweep/ejection events in the flow. In particular, it is known
that high levels of wall-shear stress (lighter zones) correspond to a sweep 
event (fast velocity directed  towards the wall) while the lower values (darker
zones) 
are  associated to the ejection events (slow velocity directed away from the wall)
\cite{pope}. 
The visualization in figure \ref{fig2}$a)$ displays the location of all particles, 
initially released both inside and outside the boundary layer. The particles
tend to preferentially localize in regions of low wall-shear i.e.\ slow ejection 
events. This is consistent with the previous results from simulations in pipe and channel flows for 
 the cases of most accumulating particles, $St^+\sim 25$ \cite{solrew,pic_pipe,sar_jfm}. 
The viscous Stokes number of the population with nominal Stokes number $St_0=30$
decreases from $St^+=24$ at $Re_\theta=2000$ to $St^+=22$ at $Re_\theta=2500$ in the region depicted in the figure. 
As observed for parallel wall-bounded flows, the particles tend to stay in elongated streaky structures also in the spatial 
developing boundary layer, although this behavior appears to be less accentuated.  
This difference can be explained by the higher value of the friction 
Reynolds number $Re_\tau\simeq 700$ in the present case, to be compared with  $Re_\tau\simeq 150-180$ of the
typical channel flow simulations such as those analyzed in References \cite{roueat,portela2002numerical,solrew,sar_jfm}.

\begin{figure}[t!]
\begin{center}
\includegraphics[width=0.45\textwidth]{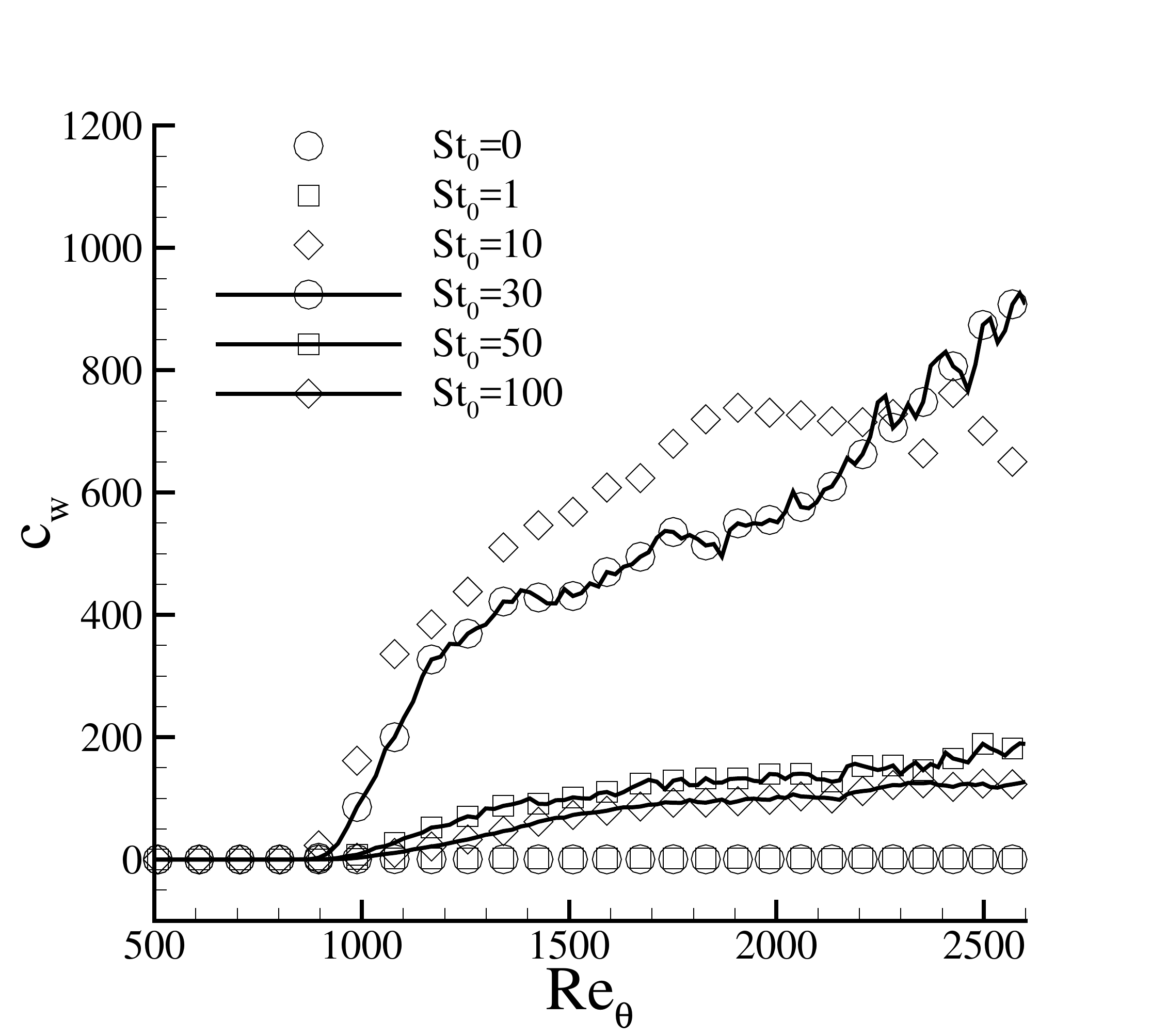}
\put(-30,110){$a)$}
\includegraphics[width=0.45\textwidth]{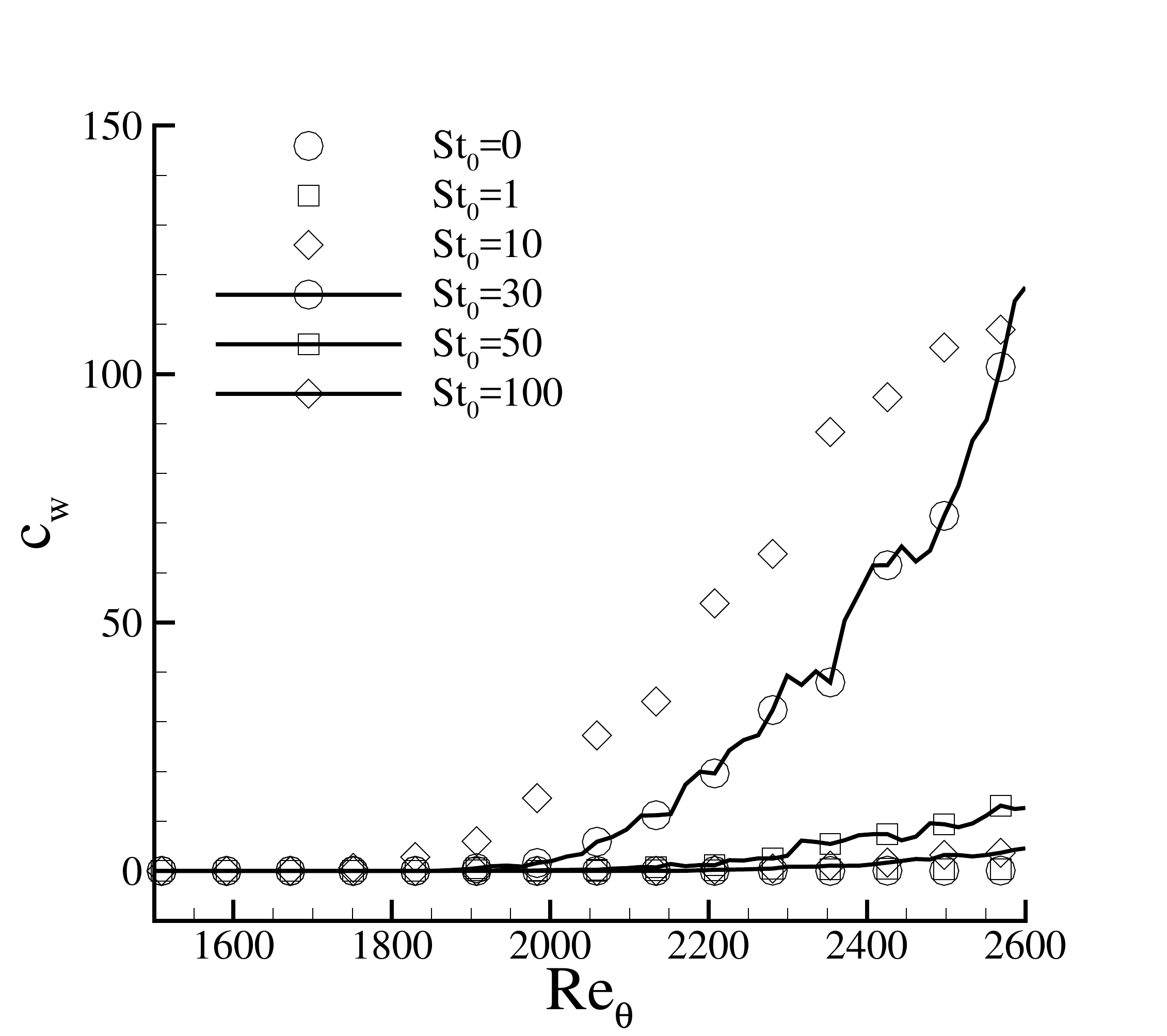}
\put(-30,110){$b)$}
\caption{Particle concentration close to the wall versus the streamwise
coordinate indicated by $Re_\theta$. $a)$ Total particles inside the domain. $b)$
Particles initially released outside the boundary layer thickness. 
 \label{fig3}}
\end{center}
\end{figure}

The particles displayed in figure \ref{fig2}$b)$ are all initially released 
 outside the geometrical thickness of the boundary layer, $\delta_{99}$.
At $Re_\theta=2000$ relatively few of these particles have reached the wall although 
 their concentration at the wall increases monotonically when moving
downstream along the domain. At the beginning of the accumulation phase
particles do not show a preferential localization in the ejection
events; only further downstream when a clear peak in the near-wall concentration profiles is forming,
particles start to preferentially sample
the low ejection events (low local wall-shear stress region). This observation
is consistent with the dynamics of the spatial evolution of the turbophoresis discussed previously
for  pipe and channel flows \cite{pic_pipe,sar_ftac}. At statistical steady state, particles accumulate in regions of vertical fluid motions away from the wall to compensate for the turbophoretic drift towards the wall, yielding in this way a zero net flux.

\subsection{Particle concentration}

The spatial evolution of the near-wall accumulation is reported in figure~\ref{fig3}. Fig.~\ref{fig3}$a$ shows the wall concentration along
the streamwise position identified by $Re_\theta$.
The wall concentration has been defined as the number of particles 
per unit volume found below a distance of $0.01 \delta_0^*$ from the wall.
The transient accumulation phase is characterized by a strong
particle drift towards the wall, i.e.\ turbophoresis, proportional to the local slope
of the concentration profiles. This accumulation phase ends at a different streamwise distance according to the particle inertia 
and in general between $Re_\theta=1100 \div 1500$. The particle populations characterized by a
nominal Stokes number of $10$ and $30$ exhibit the highest turbophoretic drift.
This phenomenon can be explained considering that the viscous Stokes number of these particles,  
$St^+=8\div30$, lies in the range of maximum turbophoresis when  $Re_\theta=1100 \div 1500$.

After this phase, we observe a secondary growth, characterized by a
less steep slope. All the populations with $St_0 > 10$  remain in
this second accumulation phase until the end of the computational domain, where
  particles with intermediate values of the relaxation time assume the highest wall concentration.
At the end of this secondary phase, a peak in the concentration  is reached only by particles with $St_0=10$,
 at the streamwise location corresponding to $Re_\theta\simeq2200$; downstream of the maximum this population shows
 a slight decrease of the wall concentration. We expect a peak in the concentration for all the population once the local Stokes number becomes of order 20; however, this would happen at a larger distance from the computational inlet and an even
longer computational domain would be required to capture this effect, something computationally too expensive for such kind of simulations.
Theoretically, in an infinitely long streamwise domain all
the particle populations will reach a concentration peak after the second
accumulation phase, with the peak position dependent
on particle inertia. This phenomenon can be explained considering that around $St^+=\sim 25$ the turbophoresis is maximum and
that in a turbulent boundary layer $St^+$ is always diminishing with the streamwise distance, i.e.\ $Re_\theta$. 
Downstream of the location of maximum accumulation, the wall concentration will diminish since particles tend to the Lagrangian limit
as their viscous Stokes number still decreases.

The near-wall accumulation of particles initially seeded outside the turbulent shear layer is displayed in
figure \ref{fig3}$b)$.
The concentration increases monotonically for the most accumulating particle families
and the behavior is similar to the phase of transient accumulation described
in the discussion of the previous plot. The accumulation process starts at $Re_\theta=2000$ and
continues until the end of the computational domain. Also in this case the
most accumulating particles are those characterized by the intermediate nominal
Stokes numbers $St_0=10,30$. Note that a clear peak for the near-wall accumulation cannot yet be distinguished in this case: the initial seeding has an effect on the maximum concentration and on the local values of the Stokes number where this maximum occurs, as the peak in concentration is reduced and moved further downstream when particles are seeded outside the boundary layer.

\begin{figure}[t!]
\begin{center}
 \includegraphics[width=0.45\textwidth]{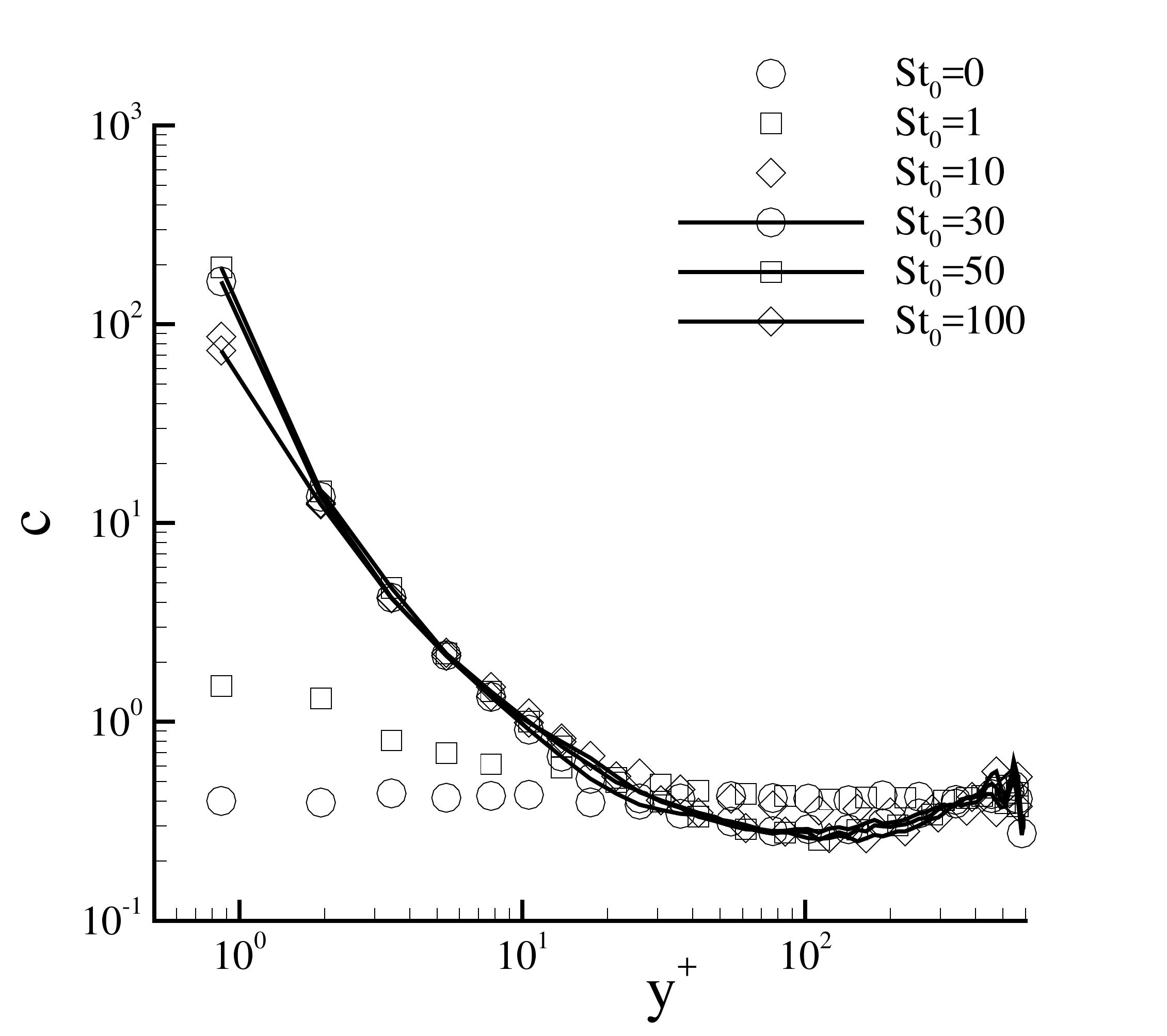}
 \put(-30,60){$a)$} ~~~
 \includegraphics[width=0.45\textwidth]{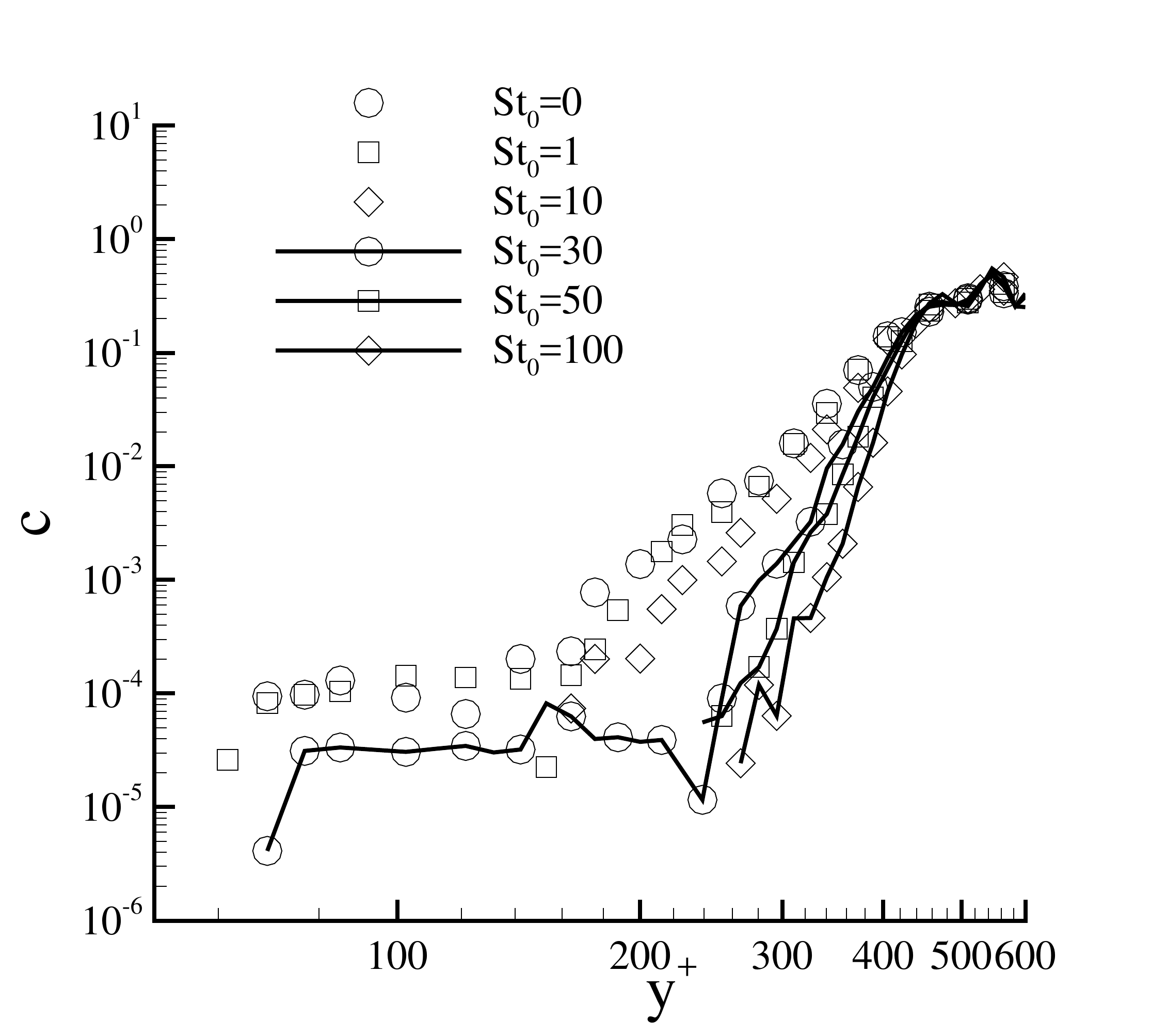}
 \put(-30,60){$b)$}
 \\
 \includegraphics[width=0.45\textwidth]{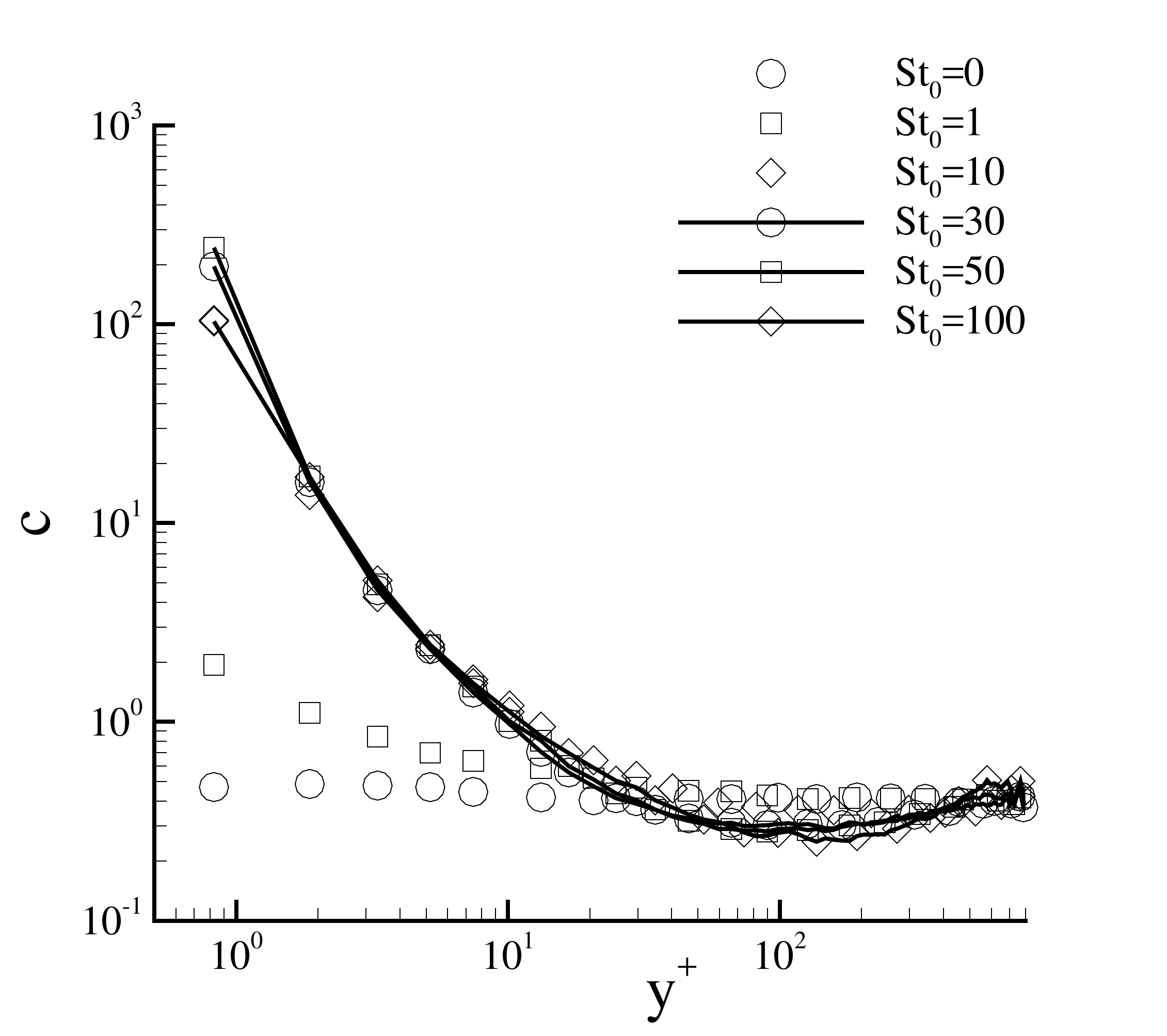}
 \put(-30,45){$c)$} ~~~
 \includegraphics[width=0.45\textwidth]{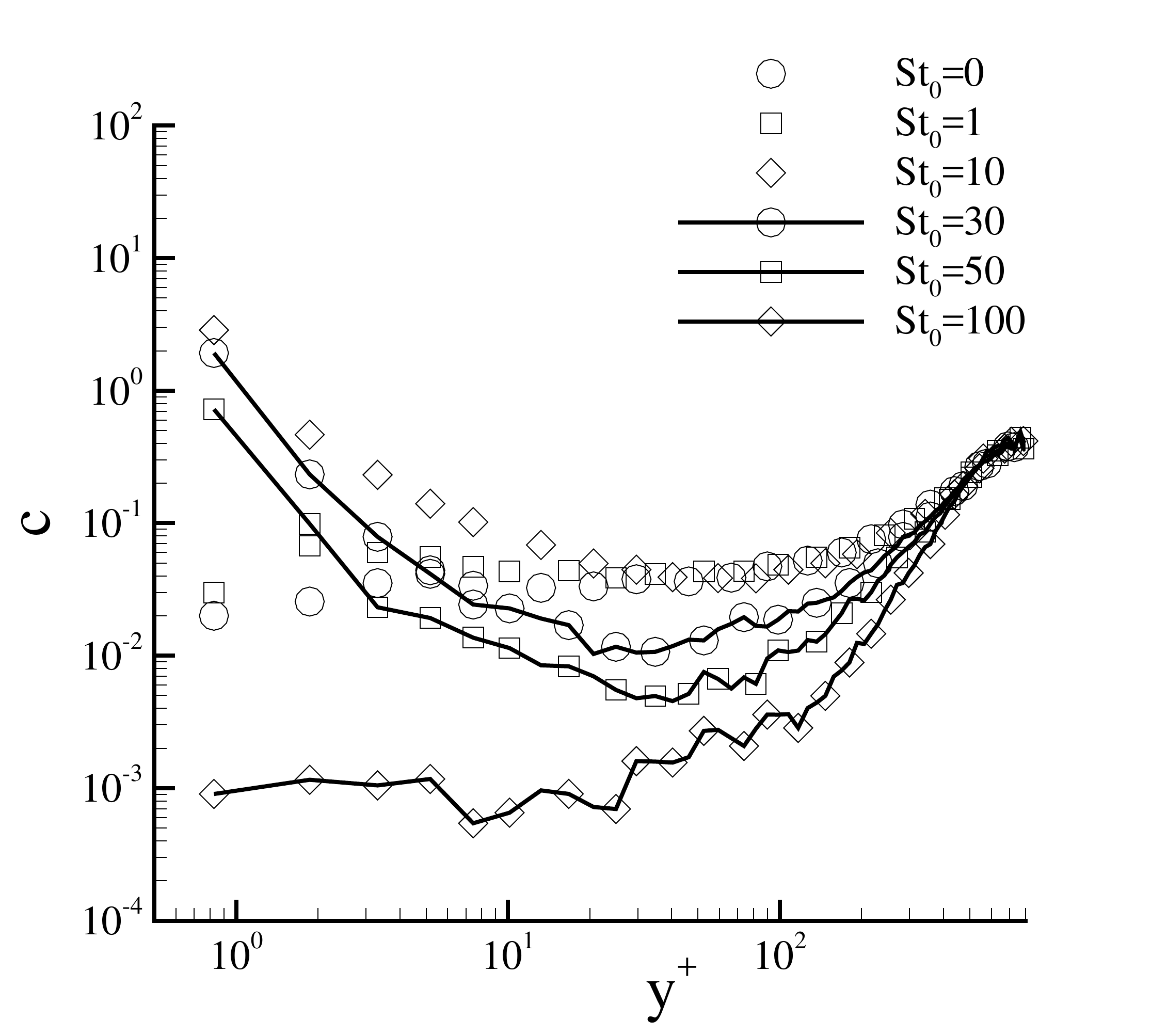}
 \put(-30,45){$d)$}
 \\
 \includegraphics[width=0.45\textwidth]{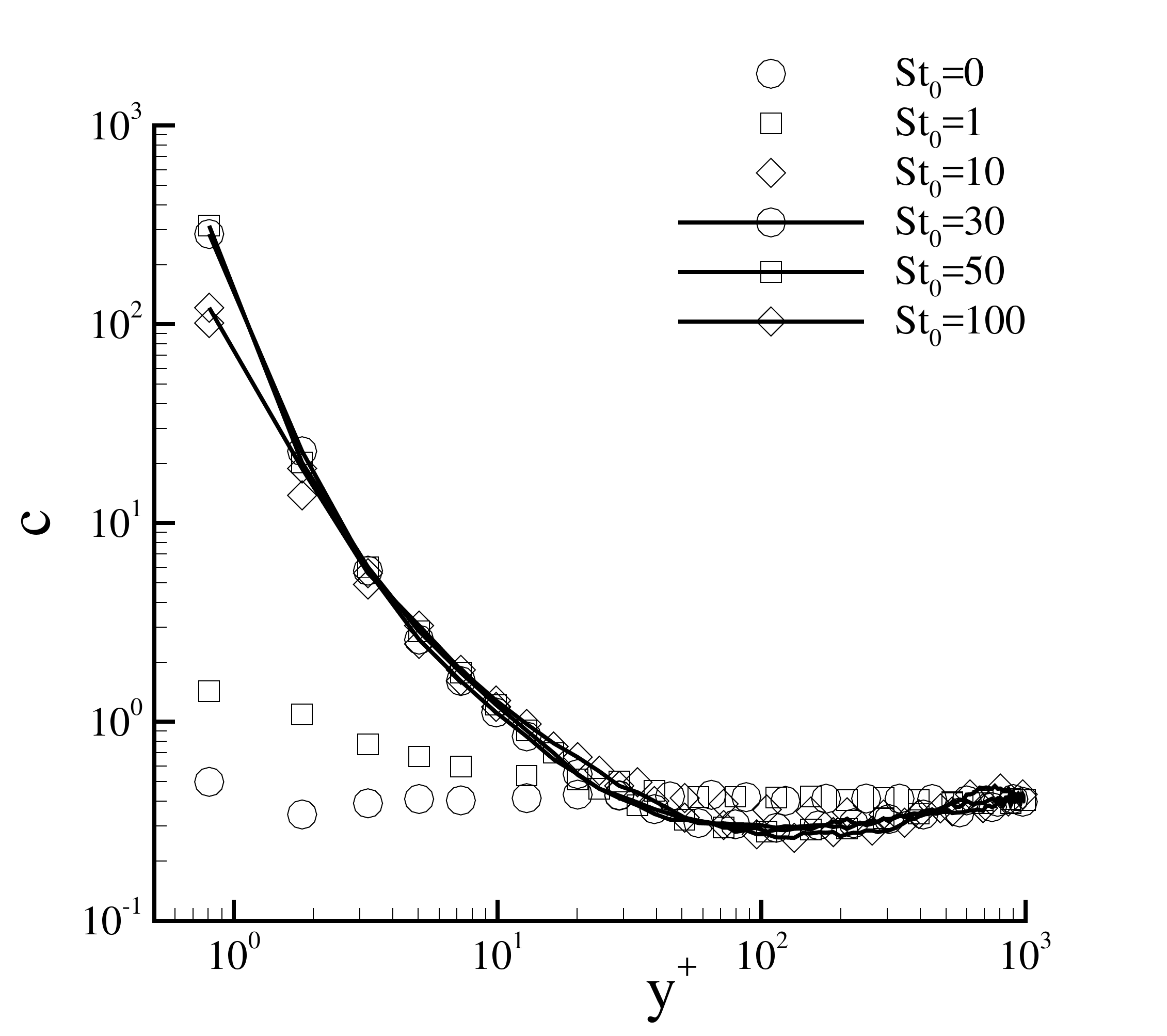}
 \put(-30,60){$e)$} ~~~
 \includegraphics[width=0.45\textwidth]{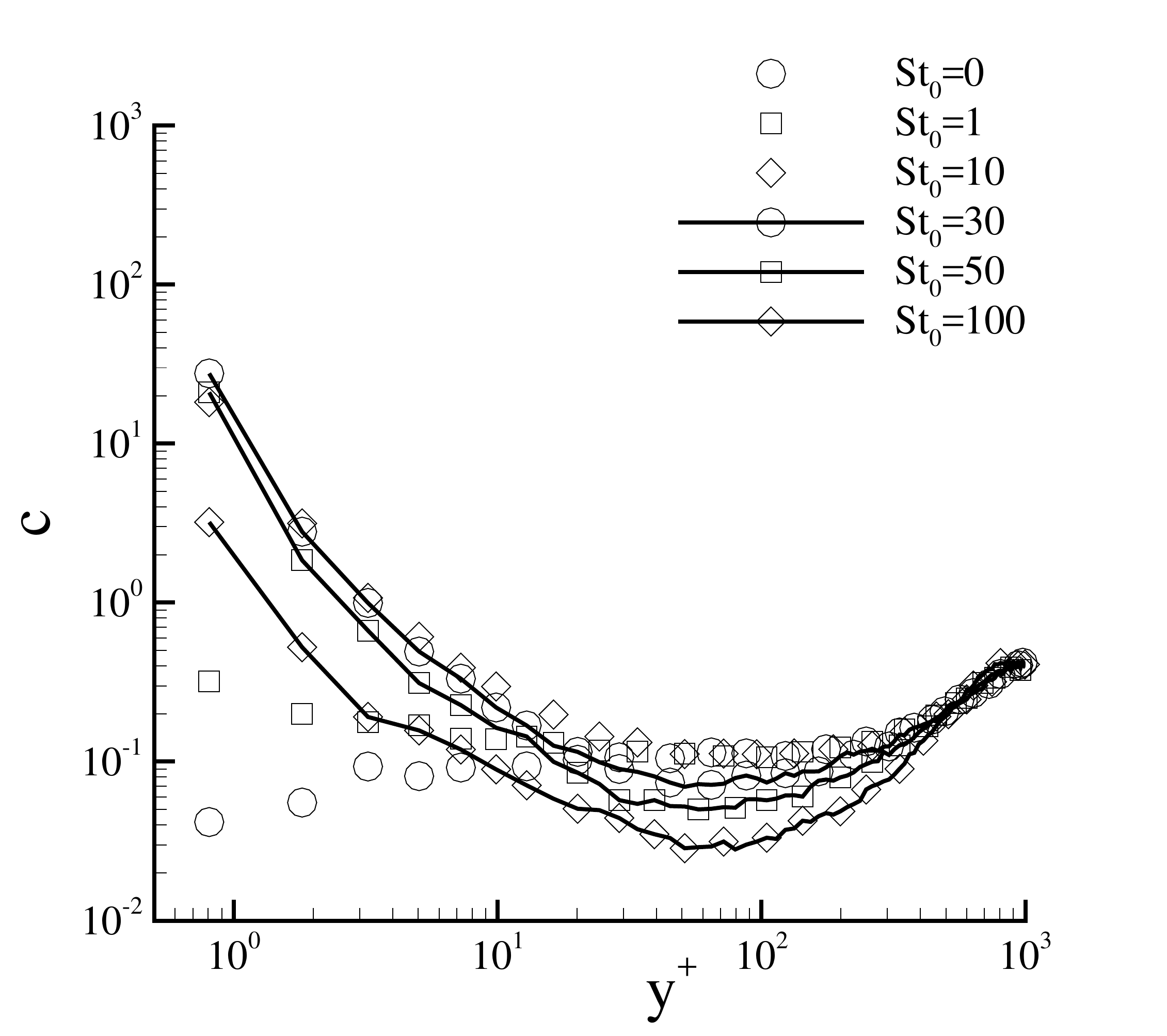}
 \put(-30,60){$f)$}
\end{center}
 \caption{Wall-normal profiles of the mean particle concentration in inner units $y^+$. Left panels: all particles inside the computational domain; right panels: particles initially released outside the boundary 
layer thickness. $a)$, $b)$ profiles at $Re_\theta=1500$. $c)$, $d)$ profiles at $Re_\theta=2000$. $e)$, $f)$ profiles at $Re_\theta=2500$. 
  \label{fig4} }
\end{figure}

\begin{figure}[tbh!]
\begin{center}
 \includegraphics[width=0.45\textwidth]{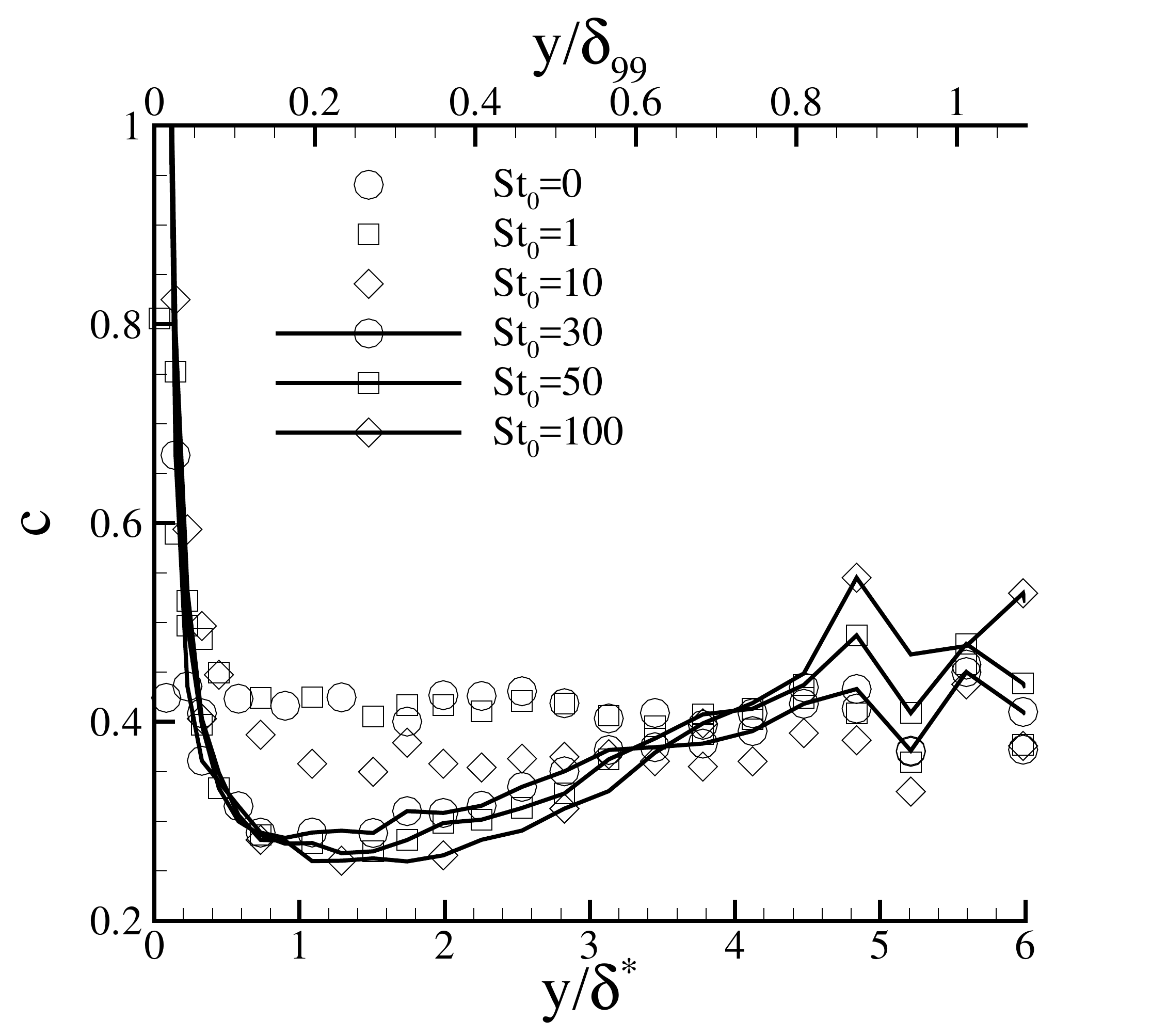}
\put(-40,90){$a)$}
~~~
 \includegraphics[width=0.45\textwidth]{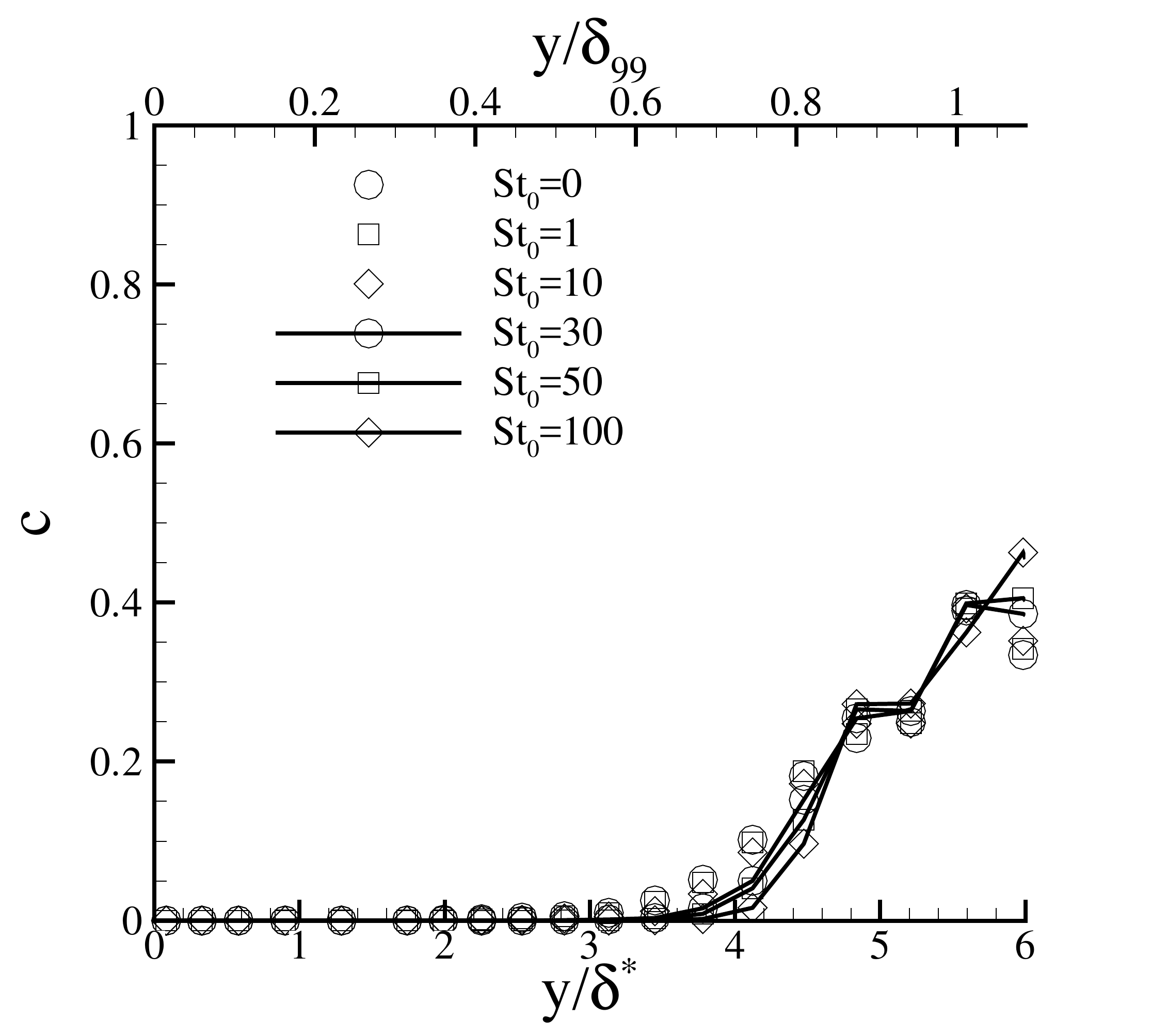}
 \put(-40,90){$b)$}\\
 \includegraphics[width=0.45\textwidth]{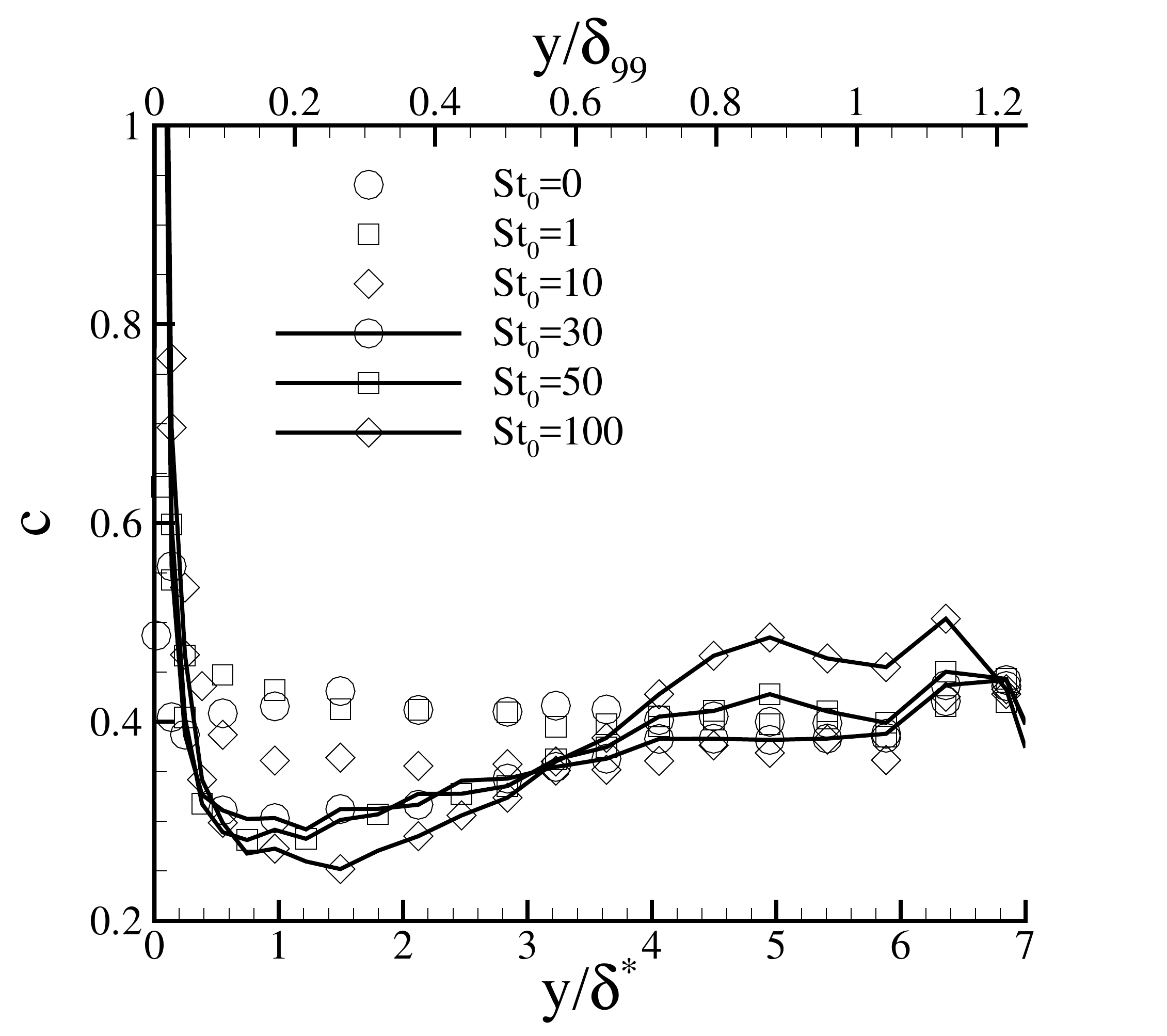}
\put(-40,90){$c)$}
~~~
 \includegraphics[width=0.45\textwidth]{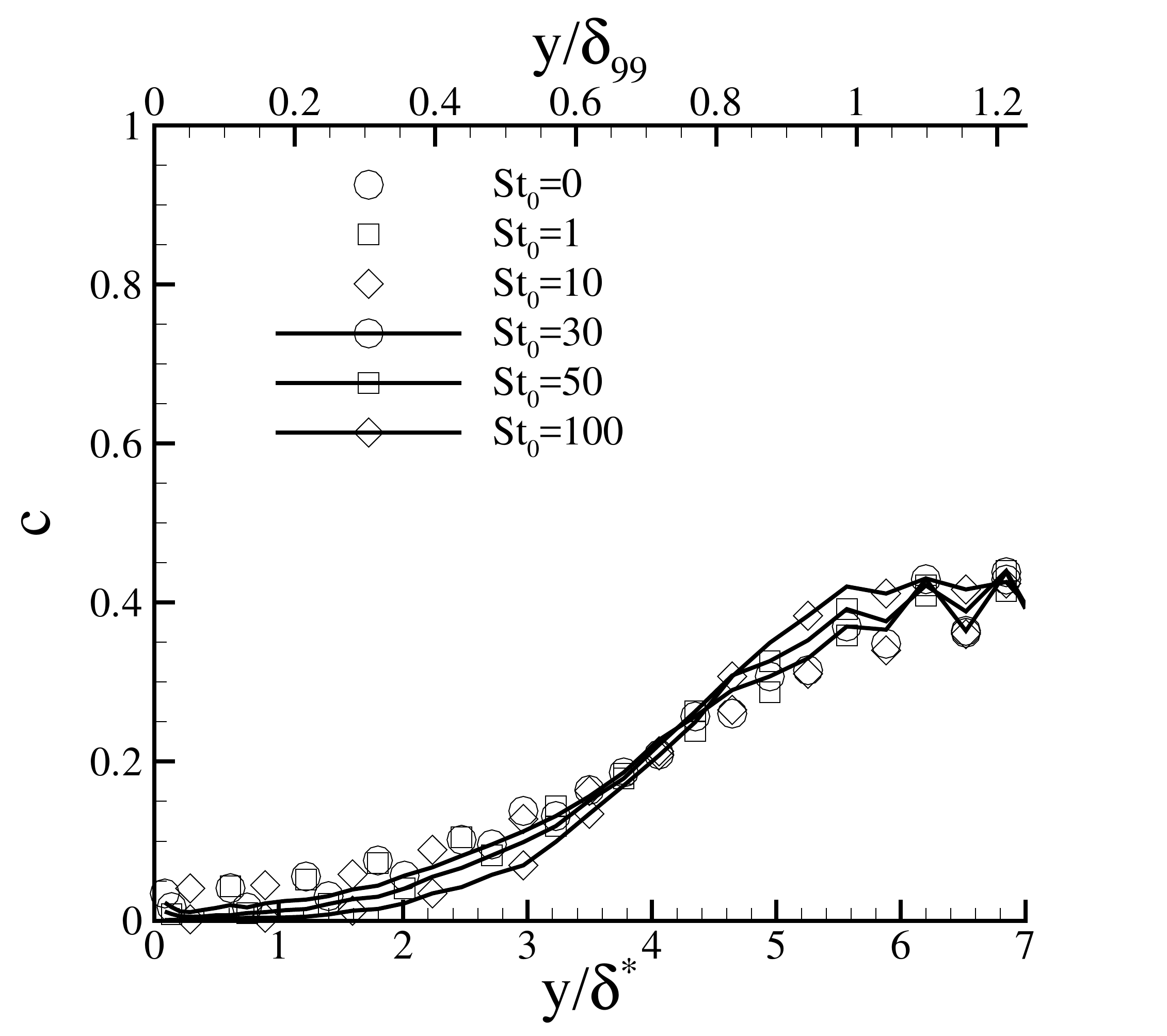}
 \put(-40,90){$d)$}\\
 \includegraphics[width=0.45\textwidth]{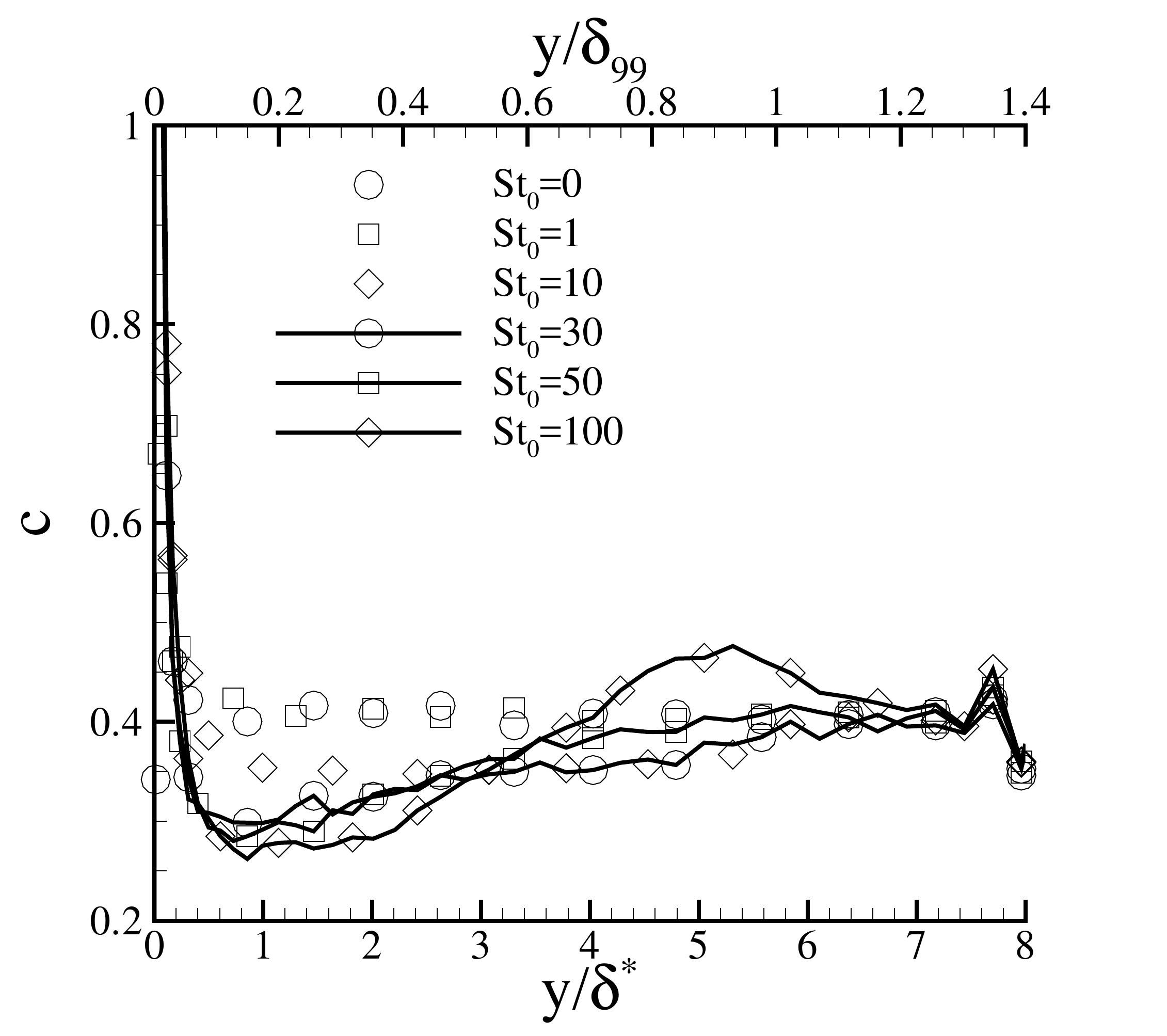}
\put(-40,90){$e)$}
~~~
 \includegraphics[width=0.45\textwidth]{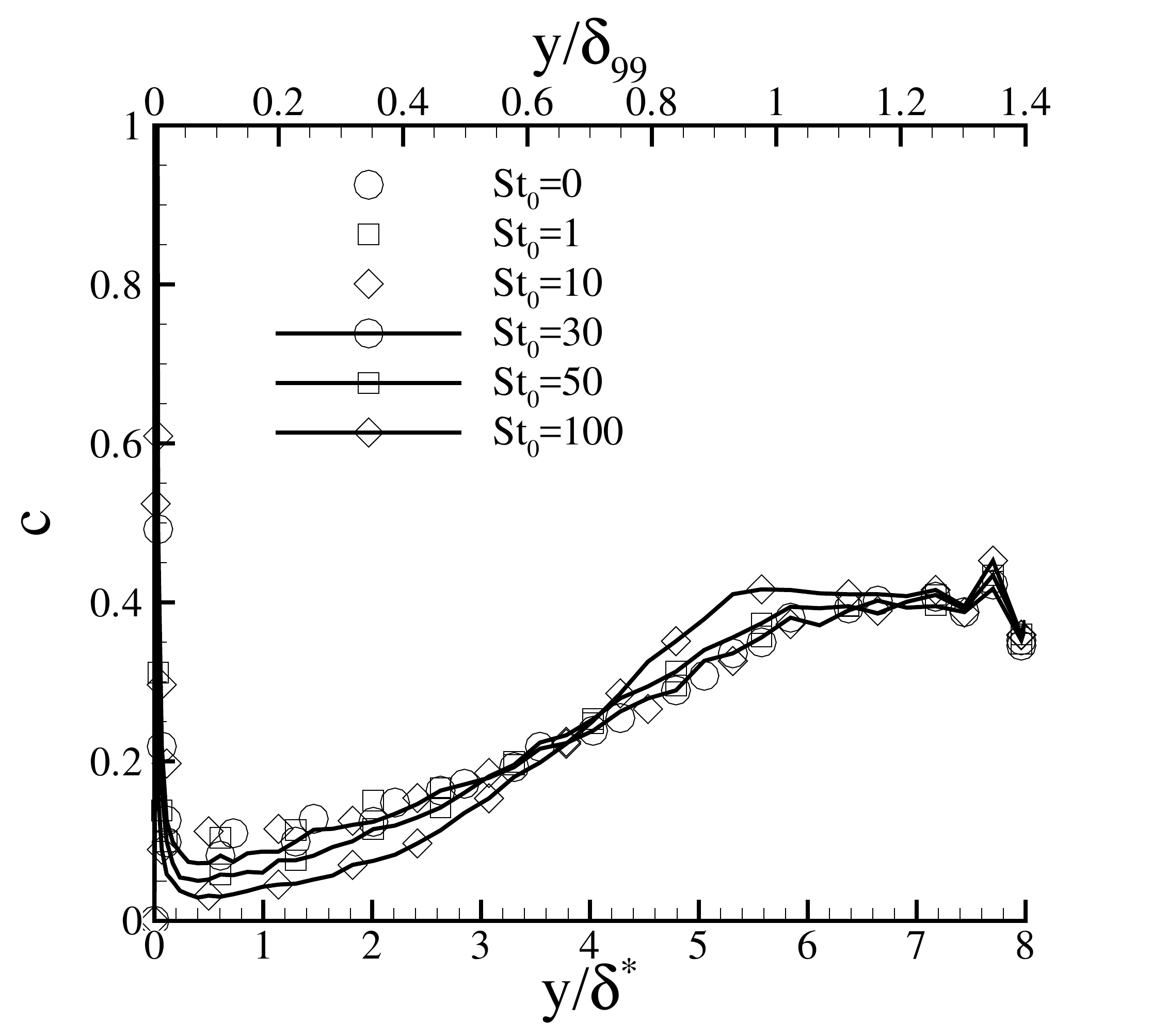}
 \put(-40,90){$f)$}
\end{center}
 \caption{Wall-normal profiles of the mean particle concentration in outer units, $y/\delta^*$ and $y/\delta_{99}$. Left panels: all particles inside 
the domain; right panels: particles initially released outside the boundary 
layer thickness. $a)$, $b)$ profiles at $Re_\theta=1500$. $c)$, $d)$ profiles at $Re_\theta=2000$. $e)$, $f)$ profiles at $Re_\theta=2500$. 
  \label{fig5} }
\end{figure}

Figure~\ref{fig4} shows the wall-normal 
concentration profiles of different particle populations, $St_0$, at the three streamwise locations corresponding to $Re_\theta=1500;\,2000;\,2500$.
On the left we show the statistics of all particles tracked in the simulation, whereas on the right we present results only for those initially released outside the turbulent boundary layer.  
Focusing on the left panels, the turbophoresis is apparent for the particles with $St_0=10\div100$ that exhibit mean wall concentrations more than
100 times that of tracers, i.e.\ $St_0=0$, with no relevant differences when changing $Re_\theta$. 
Particles of small inertia, $St_0=1$, do not show relevant turbophoresis, though still appreciable.
Unlike the case of a turbulent channel flow, we see a minimum of the concentration 
for turbophoretic particles in the outer part of the boundary layer, $y^+ \approx 100$ in the figure, before the concentration recovers the unperturbed values further away from the wall. 
This minimum is originated by the combination of the strong turbophoretic drift
that move the particles towards the wall and a gentler particle dispersion from the outer layer towards the bulk of the turbulent boundary layer. 
Indeed, the concentration profiles for the particles released in the external region of the boundary layer, plotted in the 
right panels of figure~\ref{fig4}, show at the location
corresponding to $Re_\theta=1500$ that the particles are still mainly concentrated in the outer region $y^+>300\div400$ and tend to slowly penetrate towards the wall; the near-wall concentration is however still negligible. The smaller the particle inertia the larger is the wall-normal turbulent diffusion that brings the particle from the external to the inner region.  
Moving downstream, $Re_\theta=2500$, the turbophoretic drift becomes apparent from the near-wall peak close to the wall. Particles with 
$0<St_0\le50$ start to display values of concentration at the wall higher than those in the outer flow. This process leads to a  concentration minimum 
in the outer part of the turbulent boundary layer that can be explained as follows.
The particles reach the buffer layer by turbulent diffusion and are there accelerated towards the wall by the turbophoretic drift. This creates a region partially depleted of particles where the turbulent diffusion cannot compensate for the increased drift towards the wall provided by the turbophoresis.

The location of the minimum concentration scales in outer units, $y/\delta^*$, as apparent from the left panels of figure~\ref{fig5}. The concentration profiles show the minimum at $y\simeq\delta^*$ for all accumulating particles and $Re_\theta$, while they become almost flat in the external free stream at $y>5\delta^*\simeq\delta_{99}$. 
Particles initially released in the free stream, are shown in the right panels of figure~\ref{fig5}. From the sequence in the figure one can recognize the initial uniform diffusion towards the wall, driven by the turbulent fluctuations, and finally the strong acceleration in the thin layer close to the wall, now driven by the turbophoresis, i.e.\ mainly by the particle inertia and their ability to filter high-frequency events.

The scaling of the minimum position in outer units, i.e.\ $\delta^*$, highlights that the process is essentially controlled by the dynamics of the outer part of the boundary layer. As anticipated above,  we argue that
this minimum originates from the competition between the slower turbulent diffusion of the particles in the outer region of the turbulent boundary layer and the fast turbophoretic drift.
From the right panels of figure~\ref{fig5}, depicting the mean particle concentration in outer units for the particles initially released in the free stream,
we see that the concentration is almost constant when $y>\delta_{99}$ and decreases towards the wall. Increasing $Re_\theta$, i.e.\ moving downstream, the turbulent diffusion tends to increase the concentration inside the turbulent boundary layer. At $Re_\theta=2500$, although the turbulent diffusion is still not
able to evenly mix the particle inside the turbulent region, the turbophoresis is already apparent close to the wall, see the large concentration peak for $St_0\ge10$.  These data indicate that the formation of the minimum in the concentration wall-normal profiles is due to an imbalance of the flux towards the wall driven by turbulence: a slow particle turbulent diffusion in the outer part of the turbulent boundary layer and a fast turbophoretic drift towards the wall.

\begin{figure}[t!]
\begin{center}
\includegraphics[width=0.45\textwidth]{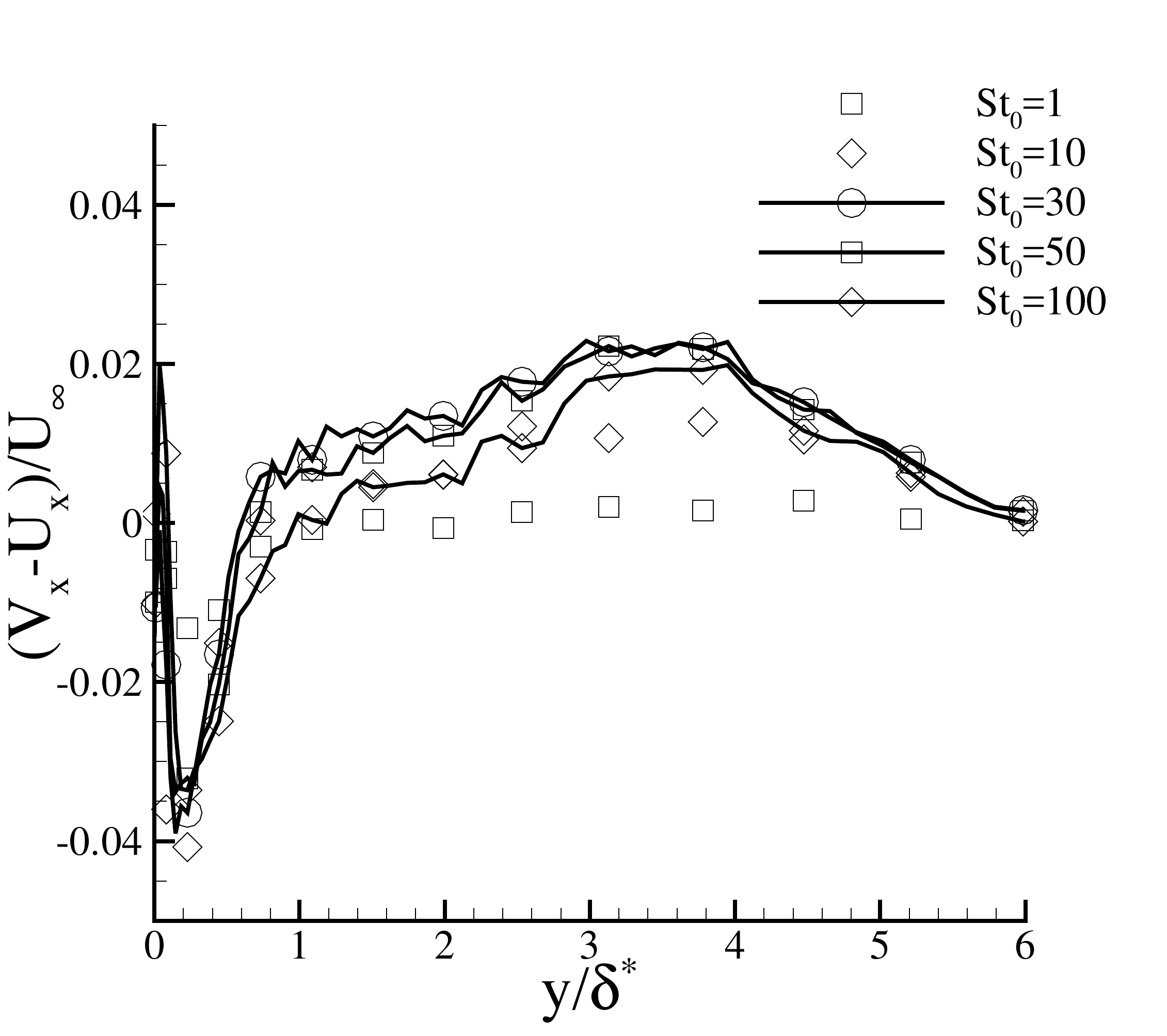}~~~
\put(-40,30){$a)$}
\includegraphics[width=0.45\textwidth]{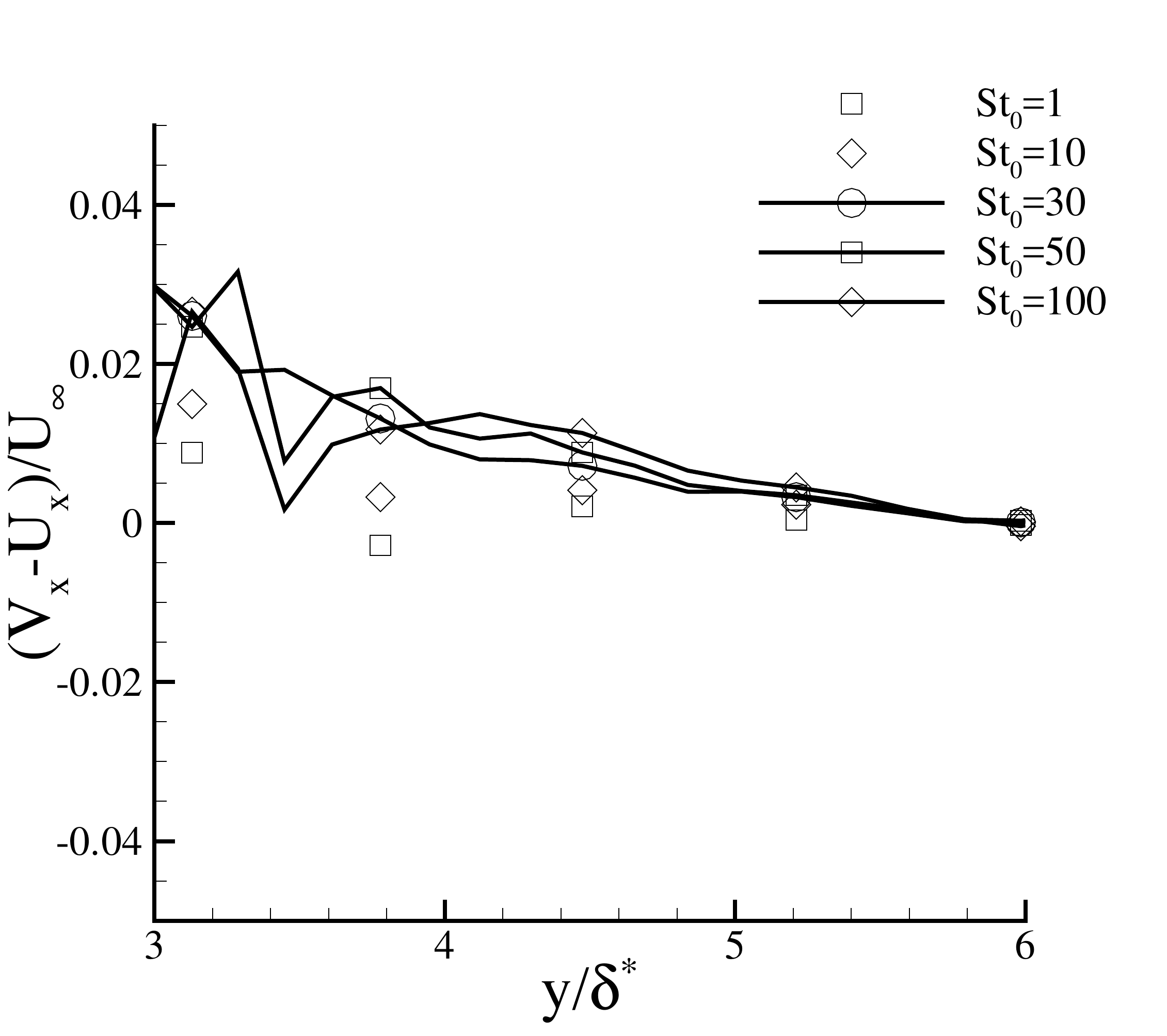}
\put(-40,30){$b)$}\\
\includegraphics[width=0.45\textwidth]{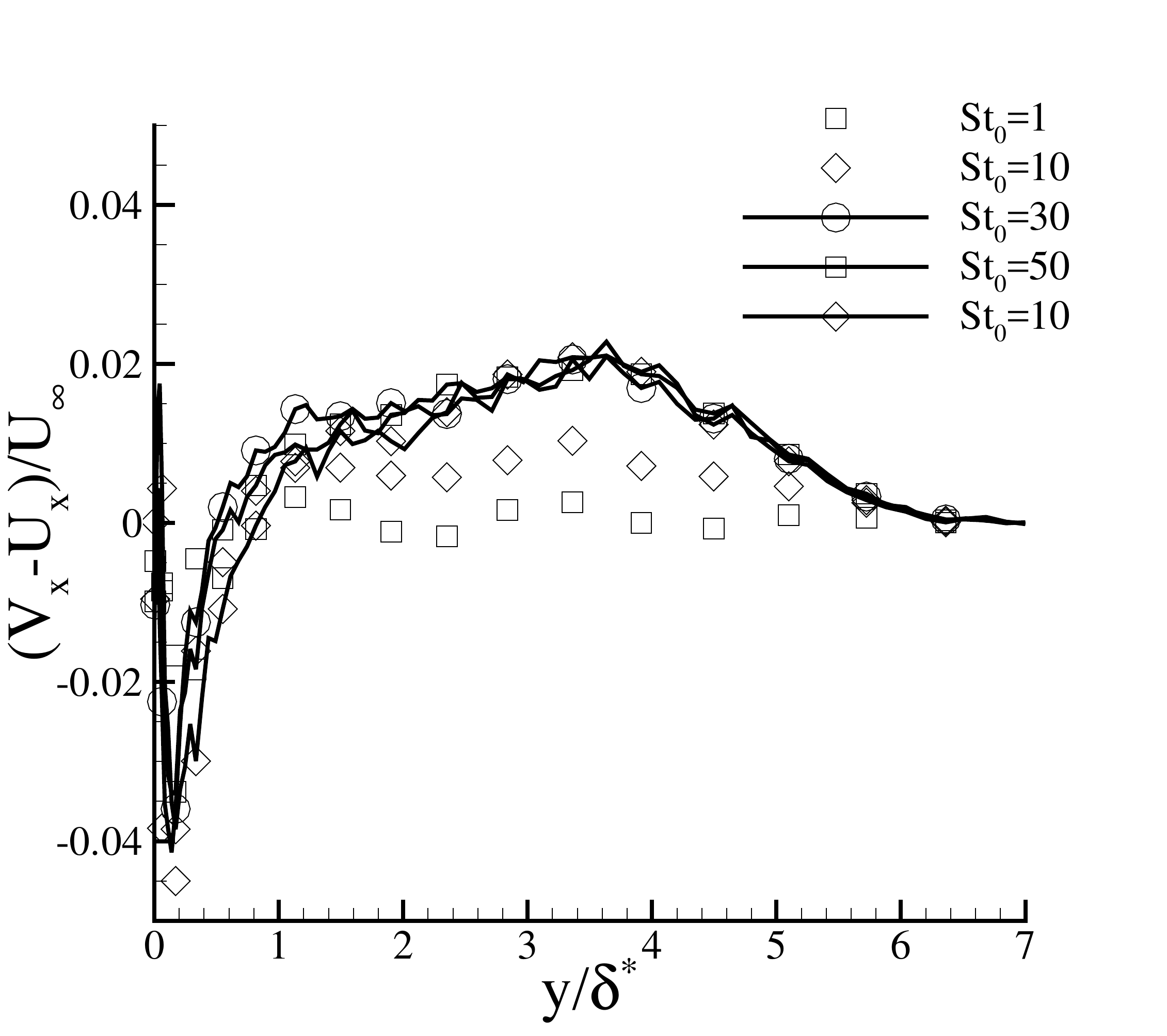}~~~
\put(-40,30){$c)$}
\includegraphics[width=0.45\textwidth]{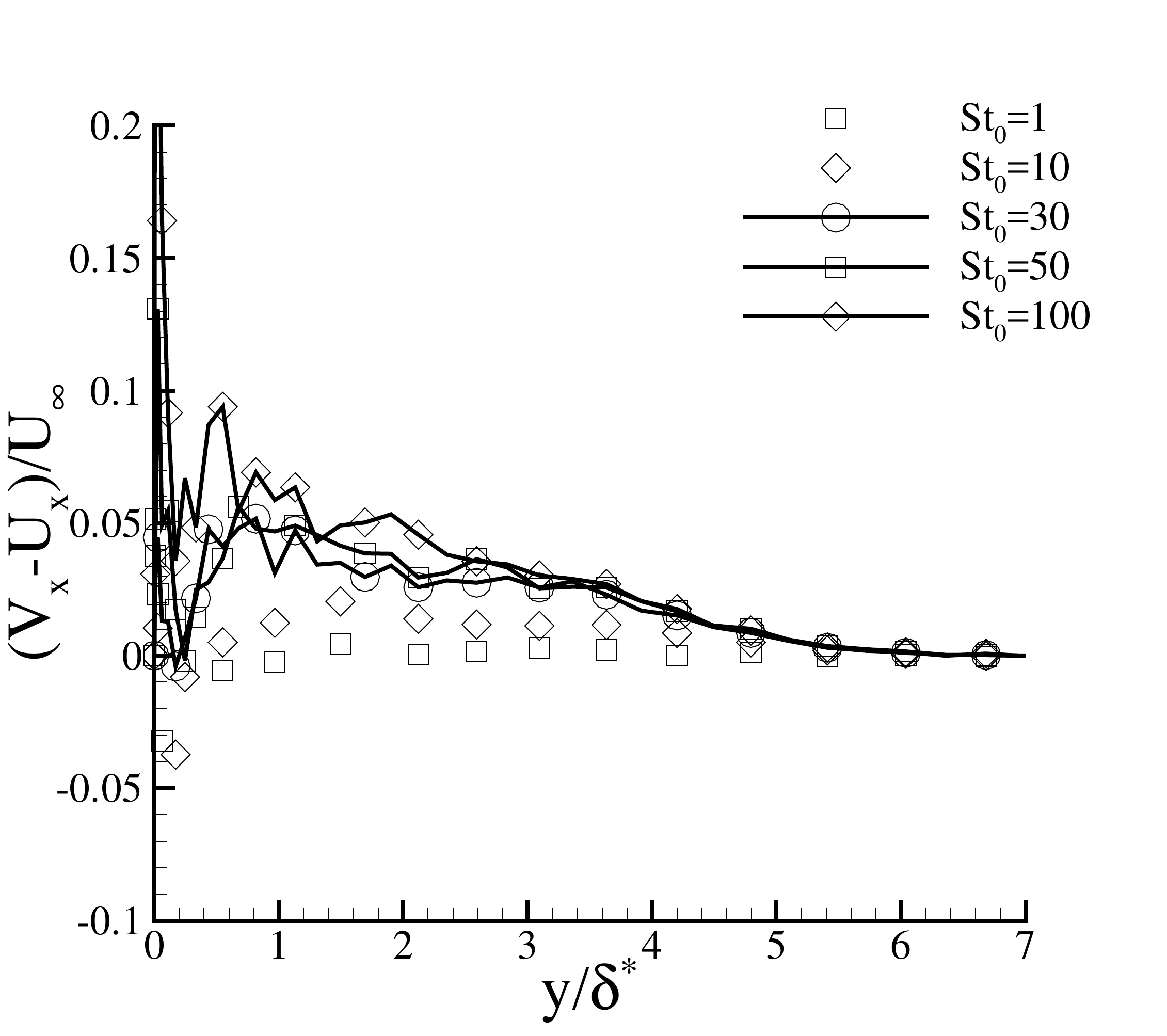}
\put(-40,30){$d)$}\\
\includegraphics[width=0.45\textwidth]{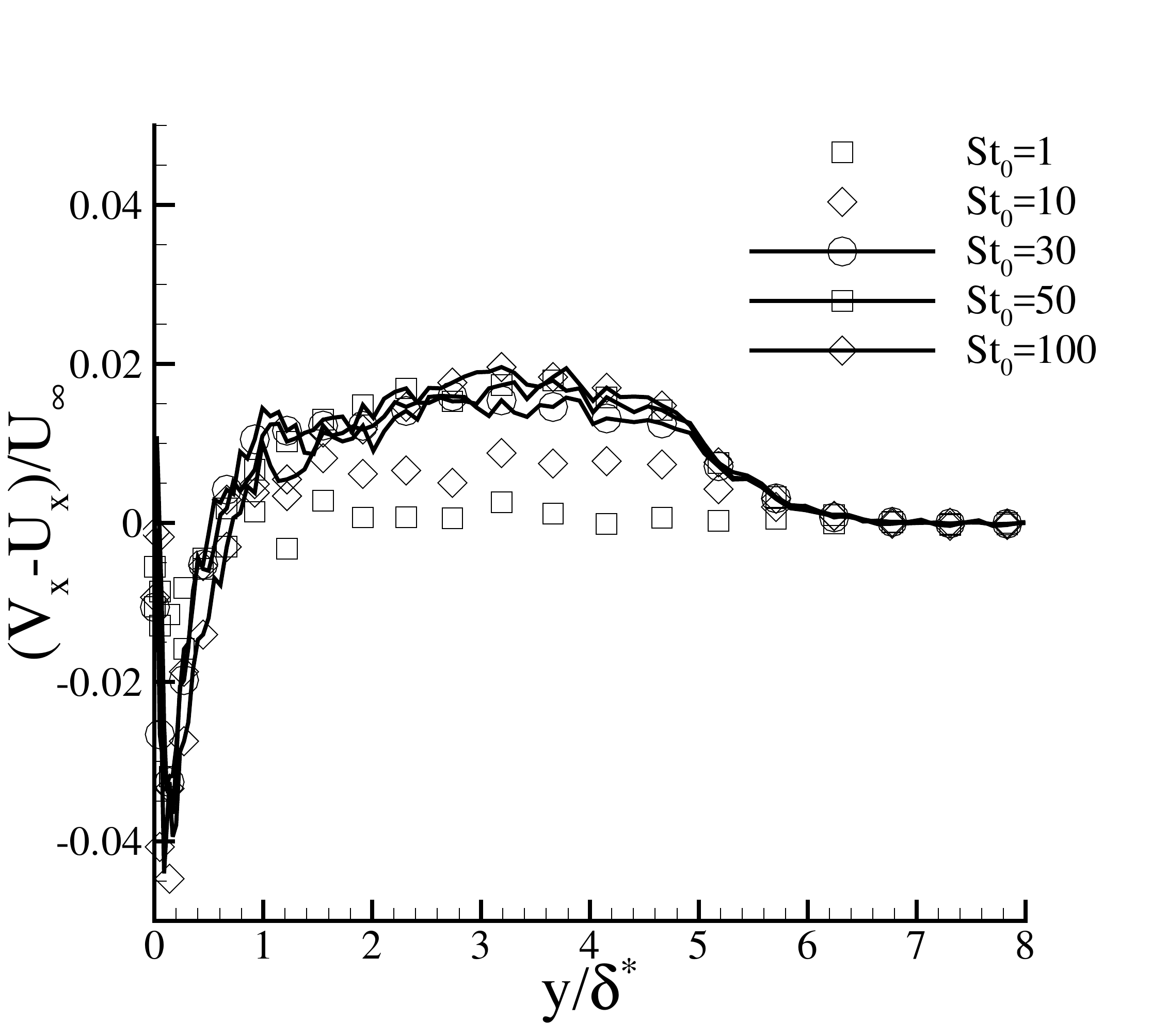}~~~
\put(-40,30){$e)$}
\includegraphics[width=0.45\textwidth]{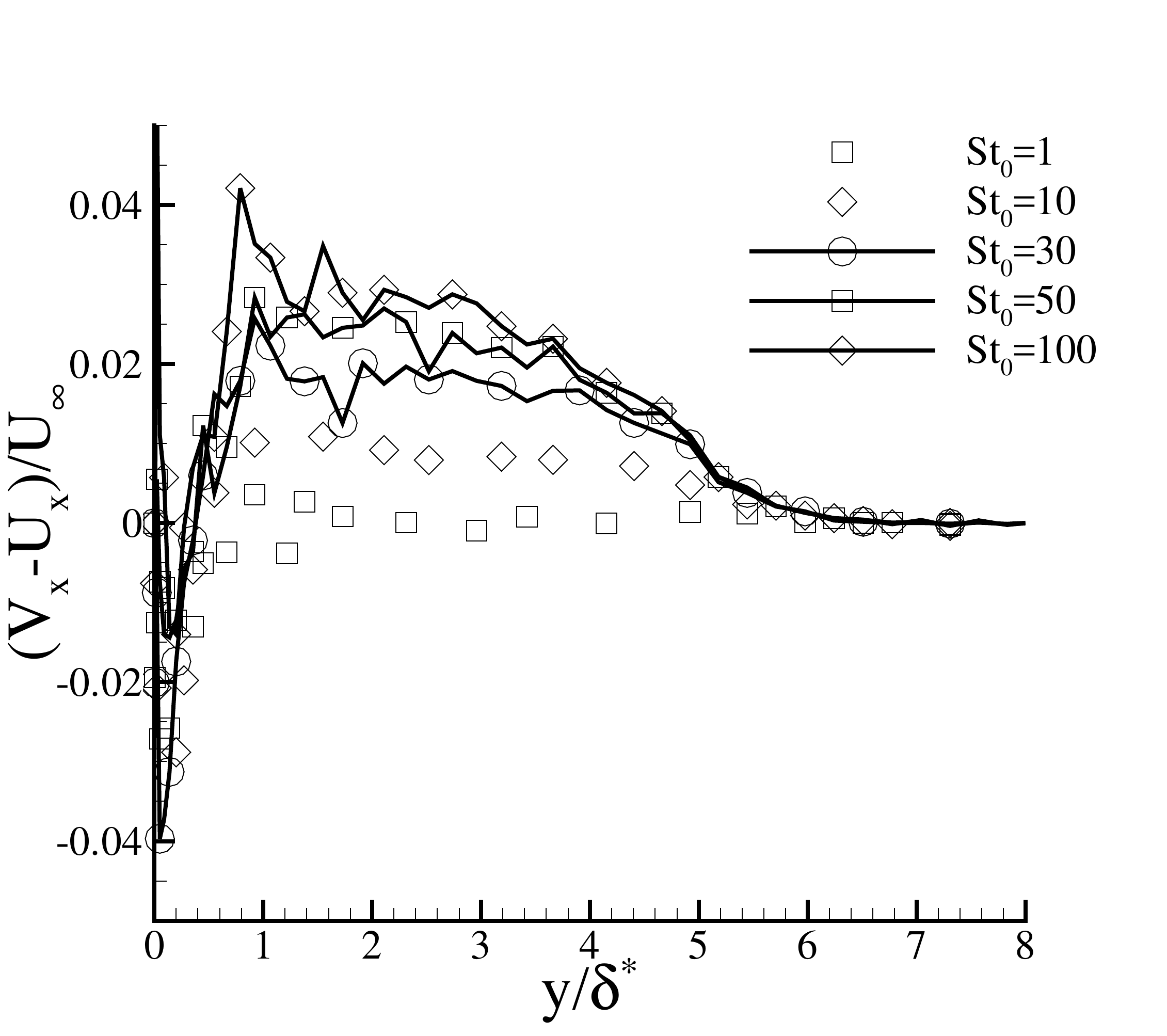}
\put(-40,30){$f)$}
\caption{Wall-normal profiles of the difference between the average streamwise
particle velocity and the mean streamwise fluid velocity. Left panels: total particles inside 
the domain; right panels: particles initially released outside the boundary 
layer thickness. $a)$, $b)$ profiles at $Re_\theta=1500$. $c)$, $d)$ profiles at $Re_\theta=2000$. $e)$, $f)$ profiles at $Re_\theta=2500$. 
 \label{fig6}}
\end{center}
\end{figure}

\subsection{Particle velocity statistics}
In order to analyze the differences between the fluid and particle motion, we report
in figure \ref{fig6} the difference between the mean particle streamwise
velocity $V_x$ and the mean streamvise fluid velocity $U_x$ 
at three streamwise positions, $a)$, $b)$ at $Re_\theta=1500$,
$c)$, $d)$ at $Re_\theta=2000$ and $e)$, $f)$ at $Re_\theta=2500$, again for all particles and for those seeded outside the boundary layer.
We first consider the global particle behavior, independently of their initial injection point
(left plots) and observe that all populations are faster than the mean flow in the outer region between $1<y/\delta^*<6$. Note
that $y/\delta^*=6$ roughly corresponds to the geometrical
thickness of the boundary layer, $\delta_{99}$. The position around $y/\delta^*=1$, in
correspondence with the minimum value of the concentration profile, is
characterized by a mean particle streamwise velocity very close to that of the fluid. 
Hence this location, one displacement thickness from the wall, can be considered an equilibrium region
for the particle dynamics. Below this point, in the region close to the
wall, particles tend to be slower than the carrier phase. This 
is linked with the preferential localization of the particles in the low-speed streaks close to the wall, as also shown from the instantaneous configurations in figure
\ref{fig4}.
Inertial particles tend to preferentially stay in the slow
ejection regions that are characterized by a streamwise velocity smaller than
the mean flow in the near-wall region as a typical feature of turbophoresis. 

In contrast with the results for the particle concentration,
the fluid-particle velocity differences do not change significantly
moving downstream.
Larger differences between the particle and fluid streamwise velocity 
 emerge for the particles initially released outside 
 the boundary layer as shown in the right panels of figure \ref{fig6}.
The data in $b)$ show the profiles at $Re_\theta=1500$ where just few particles
have entered the turbulent boundary layer. 
Here the particles tend to be faster
than the fluid as observed for  the unconditioned case reported in the figure \ref{fig6}$a)$.
Further downstream, $Re_\theta=2000$, panel $d)$, the particles coming from the outer stream 
and diffusing into the boundary layer are much faster than the fluid displaying values of the velocity difference almost twice as large as
those pertaining the unconditioned case in panel $c)$. 
Particles captured by the boundary layer and approaching 
the wall from the irrotational free stream  are characterized by a strong wall-normal velocity directed towards
the wall that is associated with a streamwise velocity larger than that of the
carrier phase.
In other words, particles are seen to enter the boundary-layer wake region via large-structures similar to the high-speed streaks of near-wall turbulence: higher streamwise velocity with wall-normal velocity towards the wall.
A similar trend is observed at  $Re_\theta=2500$, panel $f)$, even though the effect is now weakened 
because the memory of the initial seeding position tends to be lost as particles move along with the flow.

\section{Final Remarks}

We report statistics of the dynamics of inertial particles transported in a spatially developing turbulent boundary layer. 
The Reynolds number based on the  momentum thickness $Re_\theta$ increases from 200 to 2500 for the computational domain of the present simulation.

A boundary layer flow presents two main features that differentiate the behavior of inertial particles from the more studied case of a parallel turbulent flow, e.g.\ in 
channels and pipes, and it is therefore a relevant test case worth investigation. The first is the variation along the streamwise direction of the local dimensionless 
parameters defining the fluid-particle interactions. The second is the coexistence of an irrotational free-stream and a rotational near-wall turbulent flow. 

The first effect has been considered in \cite{sardina2012self}. Two different Stokes numbers have been defined, one using inner flow units
and the other with outer units. Since these two Stokes numbers exhibit different decay
rates in the streamwise direction, we found a decoupled particle dynamics between
the inner and the outer region of the boundary layer. 
Preferential near-wall particle accumulation is similar to that observed in turbulent channel flow, while different
behaviour characterizes the outer region. Here the concentration and the streamwise
velocity profiles were shown to be self-similar and to depend only on the local value
of the outer Stokes number and the rescaled wall-normal distance.

The scope of this paper is to examine the effect of the simultaneous presence of the external laminar stream and of the turbulent shear layer on the particle transport. In particular, we study how inertial particles released in the outer stream disperse inside the boundary layer and approach the wall.
To this aim, we present statistics conditioned by the position of the initial injection and discuss the emergence of a minimum of the wall-normal particle concentration around the boundary layer thickness, $y \approx \delta^*$.

The imbalance between two different diffusion mechanisms, both directed towards the wall for particle coming from the free stream, induces the 
concentration minimum in the wall-normal direction. On one side, we have a relatively slow particle dispersion in the outer part of the boundary layer, due to the mixing by 
turbulent fluctuations, bringing the solid phase from the outer region to the buffer layer. On the other side, we have a fast turbophoretic drift due to the decrease of 
turbulent fluctuations closer to the wall pushing the particles towards the wall. The different magnitude of the two fluxes creates a region of lower concentration at the 
edge of the zones where each of the two mechanisms dominate.

The entrainment of particles in the turbulent regions is occurring in regions of higher streamwise velocity and wall-normal velocity towards the wall. Note that these 
structures, though similar to the near-wall high-speed streaks, are present across the laminar-turbulent interface at the boundary-layer edge effectively leading to a 
corrugated appearance of the instantaneous laminar-turbulent interface.

A similar minimum in concentration is not observed in channel flows where a laminar region acting as a reservoir of particles and imposing a reference external 
concentration, does not exist. Hence, there is no imbalance between the two fluxes mentioned above and the minimum  concentration occurs at the channel midplane by 
symmetry.

The present results show that a non trivial dynamics occurs in intermittent regions, relevant for particle entrainment and mixing in several applications. 
Idealized shear-less configurations have been considered in \cite{war11,collins12} where both large fluctuations of the particle distribution and  self-similar 
concentration are observed.  We expect that the presence of a mean shear in an intermittently turbulent/non-turbulent  flow adds new interesting features as shown here for the entrainment of particles seeded in the laminar free stream. Using the configuration adopted here,  we therefore plan to study the behavior of inertial particles in a transitional flow where turbulent spots appear randomly in space and time \cite{brandt2004transition}.

\begin{acknowledgements}
The authors acknowledge DEISA (Distributed European Infrastructure for 
Supercomputing Applications) for the computer time granted within the project 
WALLPART. In particular, we thank EPCC (Edinburgh Parallel Computing Centre) 
for help in setting up the simulation and the assistance throughout the project. 
Resources at NSC (National 
Supercomputer Centre) at Link\"oping University,  allocated via SNIC (Swedish National Infrastructure for 
Computing), were used for post-processing the data. We would like to acknowledge 
the support from the COST Action MP0806 \emph{Particles in Turbulence}.
\end{acknowledgements}



\end{document}